\documentclass{sig-alternate-2013}
\usepackage{color}
\usepackage{url}
\usepackage{verbatim} 
\usepackage{subfigure} 
\usepackage{multirow}
\usepackage{algorithm}
\usepackage[noend]{algorithmic}
\usepackage{xspace}
\usepackage{graphicx}
\usepackage{epsfig}
\usepackage{amsmath}
\usepackage{amssymb}
\usepackage{times, tikz}
\usetikzlibrary{arrows}

\newcommand{\hide}[1]{}
\newcommand{\xhdr}[1]{\vspace{2mm}\noindent{{\bf #1.}}}

\newcommand{\ie}{\emph{i.e.}}

\graphicspath{{}}

\begin{document}

\title{The Bursty Dynamics of the Twitter Information Network}

\numberofauthors{2}
\author{
\alignauthor Seth Myers\\
\affaddr{Stanford University}\\
\email{samyers@stanford.edu}\\
\alignauthor Jure Leskovec\\
\affaddr{Stanford University}\\
\email{jure@cs.stanford.edu}\\
}

\clubpenalty=10000 
\widowpenalty = 10000

\maketitle

\begin{abstract}
In online social media systems users are not only posting, consuming, and resharing content, but also creating new and destroying existing connections in the underlying social network. While each of these two types of dynamics has individually been studied in the past, much less is known about the connection between the two.  How does user information posting and seeking behavior interact with the evolution of the underlying social network structure?

Here, we study ways in which network structure reacts to users posting and sharing content. We examine the complete dynamics of the Twitter information network, where users post and reshare information while they also create and destroy connections.
We find that the dynamics of network structure can be characterized by steady rates of change, interrupted by sudden bursts. Information diffusion in the form of cascades of post re-sharing often creates such sudden bursts of new connections, which significantly change users' local network structure.
%
These bursts transform users' networks of followers to become structurally more cohesive as well as more homogenous in terms of follower interests.
We also explore the effect of the information content on the dynamics of the network and find evidence that the appearance of new topics and real-world events can lead to significant changes in edge creations and deletions.
Lastly, we develop a model that quantifies the dynamics of the network and the occurrence of these bursts as a function of the information spreading through the network. The model can successfully predict which information diffusion events will lead to bursts in network dynamics.
\end{abstract}

\vspace{4mm}
\noindent {\bf Categories and Subject Descriptors:} H.2.8 {\bf
[Database Management]}: Database applications---{\it Data mining}

\noindent {\bf General Terms:} Algorithms; Experimentation.


\section{Introduction}
\label{sec:intro}

Online social networking and social media sites have become an ubiquitous mechanism for sharing and seeking information. In these sites, users form a network of connections by linking to friends, celebrities, organizations, and news outlets. By creating such {\em follower} connections to others, users subscribe to the content that others post. Thus, as users choose who to connect to, they are also (implicitly) choosing to which information they will have access. 


Such behavior leads to two interesting types of dynamics. First is the dynamics of creation and destruction of connections of the underlying social networks of follower relationships, while the second is the dynamics of information flow in these networks, where users produce posts that others then consume as well as {\em reshare} to their own sets of followers. 
Both of these processes are by now relatively well studied and understood. For example, dynamics of networks which evolve by creation of new links \cite{backstrom11randomwalk,yu10strucevol,jure08microevol, jure05diameter} and by the destruction of unwanted ones \cite{kwak12unfollow,xu13brokenties} has been examined. 
Similarly, study of the dynamics of creation, consumption, and resharing of information in online networks has lead to a rich body of work. Examples include predicting what content will become reshared and popular \cite{cheng14cascades, Hong2011www, kunegis2011bad, szabo2010predicting}, recommending items to others \cite{leskovec-ec06j}, quantifying the influence of users on the content consumed by others \cite{bakshy2011everyone, cha2010measuring, judd-color-2010}, and studying the propagation of pieces of information across large networks \cite{dow2013anatomy, gruhl-blogspace, kumar-conversations,leskovec-blogspace-sdm07}.

However, the interaction between these two types of dynamics is much less understood. 
%
For example, it is possible that the network could react and reconfigure itself due to the flows of information along its edges. In particular, as users in networks create and delete edges, they control the content to which they are exposed. Thus, as the information is shared from a user to user and flows through the network, users might react to it by breaking old connections and also creating new ones. For example, if the information shared by a user is offensive or not interesting, a follower might decide to drop a connection. On the other hand, if a user posts a piece of content that gets reshared through the network, others might get exposed to it and find it interesting. As a result they might decide to connect to the original poster and directly get access to the information she is posting. In both of these cases, the sharing of content affected how users are connected to each other in the network. 

It is thus important to consider the question of the {\em interaction} between the two dynamic processes: the process of users posting and sharing information, and the process of network evolution. When do information sharing events cause changes to the network dynamics?  How do these changes effect the network of a user as well as future information to which she is exposed?  Can information-driven network changes be detected and predicted?


These open questions pose a challenge because establishing the connection between the dynamics of information sharing and the dynamics of network evolution also requires understanding of how information spreads in networks. Explicit traces of information sharing and network evolution have been traditionally hard to obtain. Additionally, as large-scale information sharing events are relatively rare~\cite{goel2012structure}, it might be hard to quantify fine-grained effects of information diffusion on the underlying network dynamics. Without a richer understanding of this question, however, it is difficult to reason about networks, the mechanisms of how content spreads through them, and the connection between information and networks. 

\xhdr{Present work: Information causes bursts in network evolution}
Here, we study the dynamics of a large social network and how it is affected by users sharing information and content. 
We examine the dynamics of the Twitter follower network, where the graph is changing as users create new edges and destroy old ones. We study the complete dynamics of a subgraph of 13.1 million English-speaking users. Within this subgraph of a fixed set of Twitter users, 1.2 billion tweets were posted as well as 112.3 million new connections were formed and 39.2 million existing ones were deleted.

\xhdr{Bursts of edge creations and deletions}  We discover that the Twitter network is highly dynamic with about 9\% of all connections changing in a month. For example, an average user with 100 followers gains 10\% more followers, while also losing about 3\% of their existing followers in a given month, and overall the network is slowly densifying~\cite{jure05diameter}. There is a constant background ``flux'' of edge creation and deletion events. However, this flux gets interrupted when there is a large information cascade spreading through the network.  In particular, we find that as information gets shared through the network, it can cause abrupt changes or {\em bursts} in the dynamics of the underlying network structure. 


We discover that such information cascades result in two phenomena. First, users in a coordinated way drop their connections to the information source (we refer to this as the {\em unfollow burst}). And second, many other users almost simultaneously create new connections to the information source (we refer to this as the {\em follow burst}). Such sudden bursts in network activity can have a significant impact on a user's network structure.  We find that the similarity between a user and her followers (measured by textual similarity of users' posts) increases sharply during such bursts. For the follow bursts, the increase is caused by others discovering a user with similar interests through information diffusion and then connecting to her.  For the unfollow bursts, less similar existing followers unfollow the user, which then also results in an increased similarity of the user's followers.  Additionally, the density of connections between the user's followers also increases during a burst.  In the same manner that new followers discover the user, they also discover each other.  Overall, the bursts increase the coherence of the local network by both increasing the similarity of the connected users as well as the density of the underlying network structure.

While bursts in network dynamics are created by users resharing information, we also examine the content of tweets that cause bursts of new followers.  Using the ``Occupy Wall Street'' movement as a case study, we find evidence that external real-world events have the power to connect similar users.  As news of an event diffuses through the network, it appears that users interested in the event connect to each other to learn more about it.

\xhdr{Modeling and predicting bursts} The Twitter network dynamics comprises of a constant flux of edge creations and deletions that occasionally gets interrupted by a sudden burst in network activity. The interesting question then is whether we can model (as well as predict) whether a piece of information that gets reshared through network will result in a burst of new followers to a given user.

We develop a model that quantifies the occurrence of bursts in the dynamics of the network as a function of information diffusion.  Our model is based on the intuition that bursts of new followers occur when a user is discovered by other highly similar users who then connect to her. In this case the diffusion of information facilitates the exposure of similar other users to the target user and gives others an opportunity to link to her.  Our model quantifies the similarity of potential new followers exposed to the user's post. The model compares the similarity of potential new followers with that of others who regularly get exposed to the user's posts, and using it as a signal predicts whether a new follower burst will occur.


With our model, it is possible to make predictions about the future evolution of the network, to identify users who are about to gain many more followers, and to predict the affect of a new information diffusion event on the local network properties. 

The rest of the paper is structured as follows. In Section \ref{sec:empirical} we describe our dataset and empirically study the connection between information cascades and network evolution. Section \ref{sec:model} then proposes a model capable of predicting whether a piece of information will result in a burst of new followers. We briefly review related work in Section \ref{sec:related} and conclude in Section \ref{sec:conclusion}.

\section{Analysis of Bursts}
\label{sec:empirical}
We begin our investigations with an empirical analysis of the dynamics of the Twitter follower graph. We study how the formation of new edges ({\em follows}) and the removal of existing edges ({\em unfollows}) can be modeled as an effect of the tweets that propagate over the Twitter network.

\subsection{Dataset description}
Our dataset consists of a subset the Twitter follower graph during the month of November 2011.  We focus on English speaking users that tweeted at least once during the month. Overall, this gives us a subgraph of about 13.1 million nodes (users) with 1.7 billion follower edges. Moreover, for every edge in the network we also obtained the exact timestamp of its creation/deletion, which allows us to investigate fine-grained network dynamics that might be a result of information flows. 

Users of Twitter also create and reshare posts by {\em retweeting} them. Such resharing behavior results in information cascades, as a single post can propagate between a large number of users in the network. Thus, for every user in this network, we also analyze her complete tweeting history and reconstruct the information flows.  In total, the users of our subgraph posted 1.2 billion tweets and retweeted each other 116.3 million times.



\subsection{Twitter graph is highly dynamic}
Examining the evolution of the Twitter follower graph we find that the network is highly dynamic. Amongst our 13 million users we identified 112.3 million new follows, as well as 39.2 million unfollows. In relation to the edges that existed at the beginning of the month, nearly 7\% new edges were added, and 2.3\% of edges got removed.  Thus, even though we are observing a fixed subpopulation of Twitter users, 9\% of the edges change in a given month. This shows that the Twitter graph is highly dynamic and thus should not be thought of as an ``only-growing'' network (a network that evolves mostly by users only adding edges~\cite{jure08microevol}). In contrast, in Twitter we see approximately 1 edge deletion for every 3 edge creations. Thus, about one quarter of all network evolution events are in fact edge deletions, which means the network structure is highly fluid and dynamic.

\begin{figure}[t]
\centering
{\includegraphics[width=.48\textwidth]{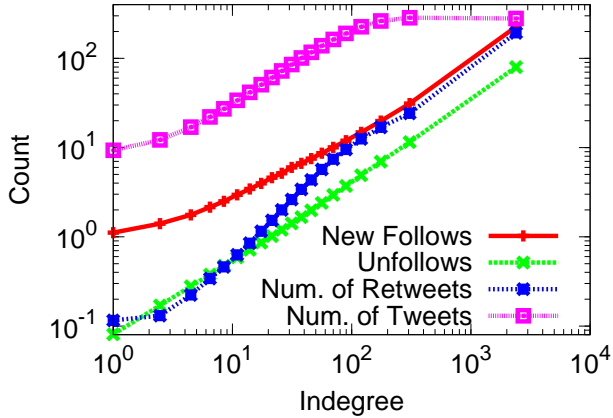}}
\caption{The number of new follows gained, follows lost, the number of retweets, and the number of tweets all scale with the indegree (number of followers) of a user. In a given month a user of degree 100 tends to gain 10 and loose 3 followers.}
\label{fig:volume_by_degree}
\vspace{2mm}
\end{figure}

To better illustrate the amount of dynamics in the Twitter network, Figure~\ref{fig:volume_by_degree} plots the average monthly activity as a function of a user's indegree (i.e., follower count) at the beginning of the month. We plot the average number of new follows, unfollows, tweets, and retweets a user receives as a function of her indegree. We observe that the average number of new follows and unfollows strictly increases with the degree of the user.  
Note that even users with only 100 followers gain on average 10 more followers during the month, while losing about 3 of them.  This high churn rate in users' followers remains consistent, even for users with high indegree. 
There is a ``background'' churn of followers, where a user is constantly gaining and losing followers.

Lastly, we also note that the distribution of new followers that Twitter users receive is heavily skewed, and as a result the follow/unfollow dynamics are heterogeneously distributed as well.  In fact, the top 20\% of users with the highest indegree receive 59.4\% of all the follows and unfollows in a given month.

\subsection{Information diffusion and follows/unfollows} 

Having observed the highly dynamic nature of the Twitter follower graph, we next focus on examining information diffusion mechanisms that might cause a user to follow someone, and also the mechanisms that may cause a user to unfollow one of her existing connections.

\begin{figure}[t]
\centering
\begin{tikzpicture}[scale=.95]

\draw (0,0) node(){$k$}circle (.5);

\draw (-1.5,-1.5) node(){$j$}circle (.5);
\draw (0,-3) node(){$i$}circle (.5);

\draw [->, ultra thick] (-1.15, -1.15)--(-.35,-.35) ;
\draw [->, ultra thick] (-.35, -2.65) -- (-1.15,-1.85) ;

\draw [->, ultra thick, red] (-.5,-.2)--(-1.3, -1);

\draw [->, ultra thick, red]  (-1.3,-2) -- (-.5, -2.8);

\node[red] at ( -1.5, -.5){ Tweet};
\node[red] at ( -1.5, -2.6){ Retweet};

\draw (4.5,0) node(){$k$}circle (.5);

\draw (3, -1.5) node(){$j$}circle (.5);
\draw (4.5, -3) node(){$i$}circle (.5);

\draw [->, ultra thick] (3.35, -1.15)--(4.15,-.35) ;
\draw [->, ultra thick] (4.15, -2.65) -- (3.35,-1.85) ;
\draw [->, ultra thick] (4.5, -2.5) -- (4.5,-.5) ;

\draw [->, ultra thick] (1, -1.5)--(2,-1.5) ;
\draw [->, ultra thick] (1, -1.2)--(2,-1.2) ;
\draw [->, ultra thick] (1, -1.8)--(2,-1.8) ;

\end{tikzpicture}
\caption{ Information diffusion and network dynamics. User $i$ following $j$ who follows $k$.  When user $k$ posts a tweet that $j$ subsequently retweets (left), then user $i$ gets exposed to the tweet, and as a result decides to follow $k$ (right). }
\label{fig:sample_follow}
\label{fig:rt_follow_example}
\end{figure}
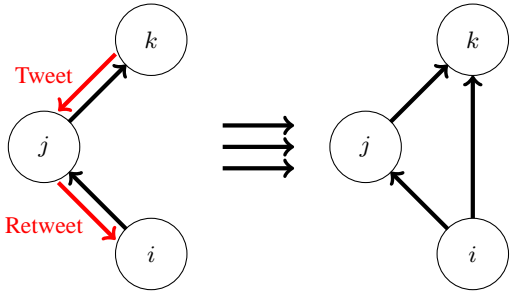

Figure~\ref{fig:rt_follow_example} illustrates an example of a process by which a new follow edge is created when a user discovers another user through a retweet~\cite{antoniades13coevol,weng13infonetevol}.  Consider user $i$ following user $j$ that follows user $k$.  User $i$ might enjoy $k$'s tweets, but $i$ does not know about $k$ and is not following her.  If $j$ happens to retweet $k$'s post, then $i$ gets exposed to it and thus learns about $k$'s existence. As a result of such exposure $i$ might decide to follow $k$. Thus, as the tweet propagates through a network, users might decide to follow the tweet originator $k$, and this way the newly created links will point up the information diffusion cascade. In fact, in our dataset 21\% of all new follows are formed by users who recently saw a retweet of the user they are newly following.  

The process that causes unfollows is somewhat different and more local. Here, current followers of user $k$ can decide to drop their connections. For example, posting offensive tweets or suddenly increasing the frequency of tweets causes users to lose followers.  This was explained by the fact that a follower who sees a particular user's tweets dominating her timeline, becomes annoyed, and then unfollows the user~\cite{kwak12unfollow,xu13brokenties}.

With these intuitions in mind, we shall now investigate whether such phenomena indeed occur in the Twitter graph.

\begin{figure}[t]
\centering
\subfigure[New follows vs. retweets]{\includegraphics[width=.23\textwidth]{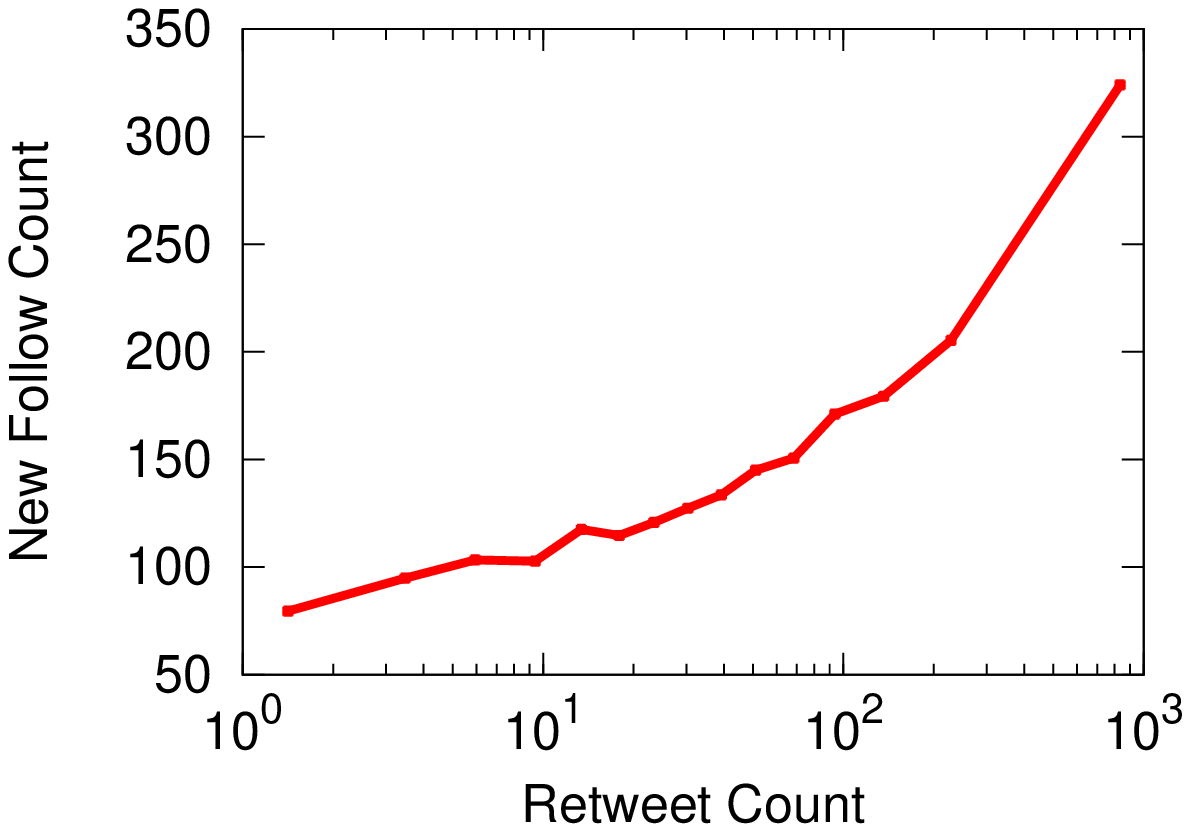}\label{subfig:rt_vs_follow}}
\subfigure[Lost follows vs. tweets]{\includegraphics[width=.23\textwidth]{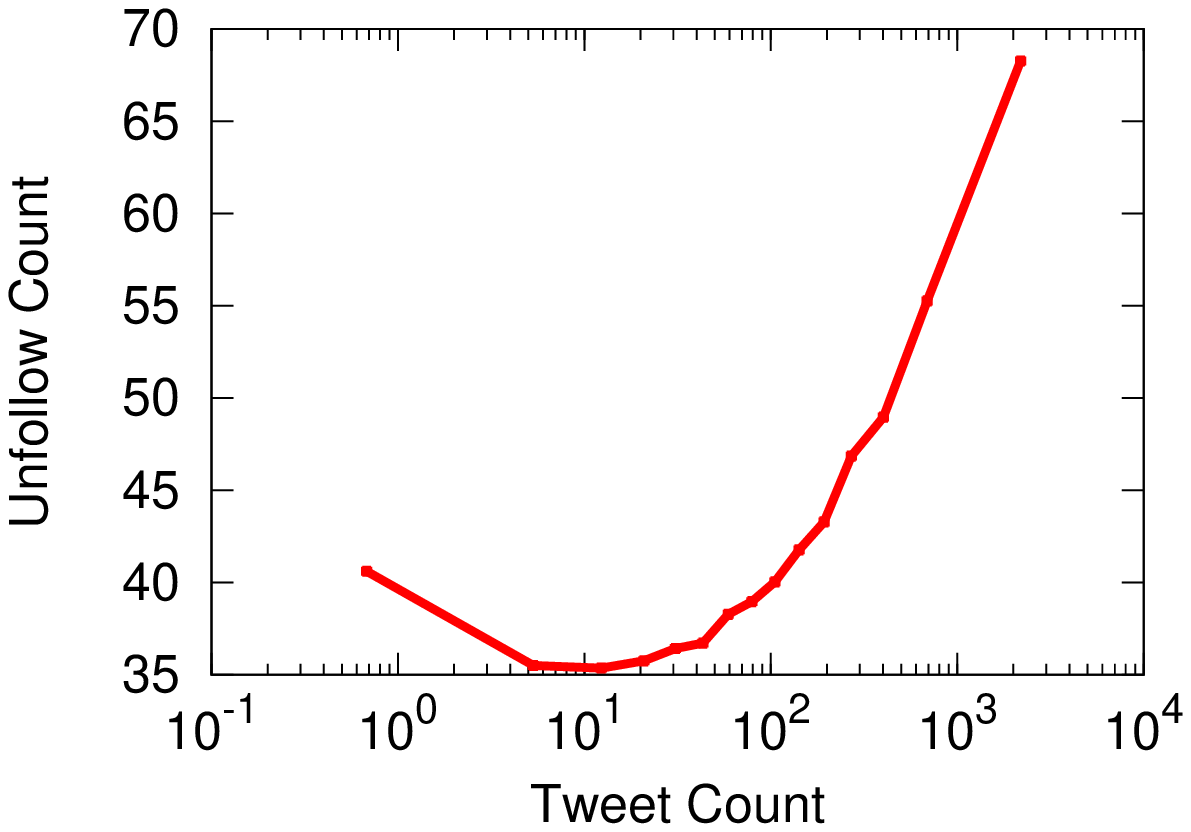}\label{subfig:tw_vs_unfollow}}
\caption{(a) The number of retweets a user acquires against the number new followers they gain, for users of fixed indegree of 1000-2000.  Even when conditioning on the indegree of the user there is a relationship between the number of retweets and the number of new followers gained.  (b) Number of unfollows as a function of tweeting activity.  Tweeting about 10 times in a month minimizes the number of followers lost.}
\end{figure}

First, we examine how a user's activity changes with her indegree. In Figure~\ref{fig:volume_by_degree} we also plot the number of tweets and retweets per user as a function of her indegree. As was the case with the dynamics of follows and unfollows, the information posting activity also scales with a user's indegree.  This could partially be explained by the fact that active users who tweet frequently tend to have more followers, and thus high indegree users get retweeted more often. 
However, even if we condition on a user's indegree, we still observe a strong relationship between users' information posting activity and the dynamics of network edges.

Figure~\ref{subfig:rt_vs_follow} plots the number of new followers as a function of the number retweets of a user's posts. Here we only consider users with indegree between 1000 and 2000. Even without the variation in indegree, there is a clear relationship between information diffusion and network dynamics.  The more retweets a user receives, the more often new users are exposed to her tweets, and the more opportunities they have to follow her.

Similarly, Figure~\ref{subfig:tw_vs_unfollow} shows the number of unfollows a user receives as a function of her tweeting activity. Interestingly, we observe a non-monotonic relationship where users who do not tweet enough, or users who tweet too much, tend to lose more followers. We find that for users of degree 1000-2000, tweeting about 10 times in a month minimizes the number of followers they loose.  

As a cautionary note, this is not to say information flow drives all of the network dynamics. As we have shown in the previous section, users also experience a steady flow of follow/unfollow events. 
Thus, we can think of the graph as being in a steady state flux, and information flow then causes perturbations to this steady state.

\subsection{Detecting bursts in the flow of followers}

\begin{figure*}[t]
    \centering
    \subfigure[user with $d_{in}=266,842$]{\includegraphics[width=0.32\textwidth]{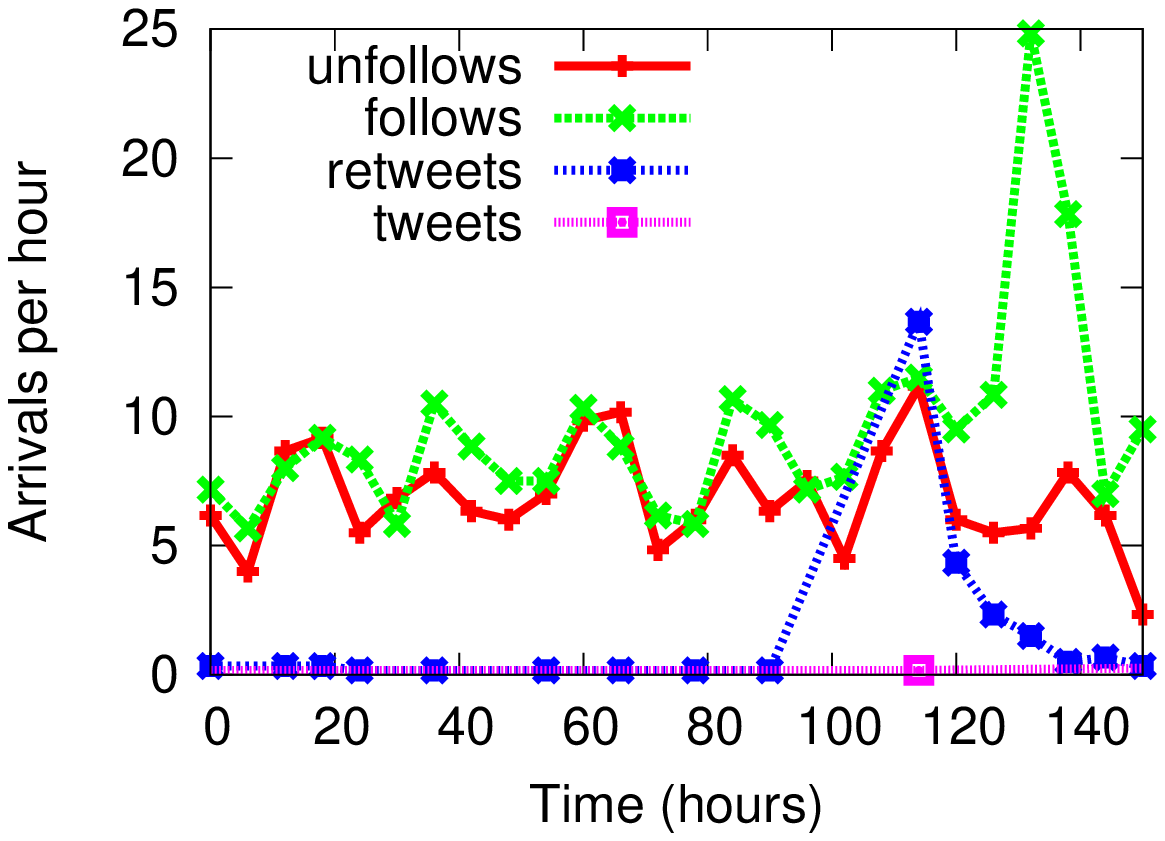}\label{subfig:u_1}}
    \subfigure[user with $d_{in}=218,045$]{\includegraphics[width=0.32\textwidth]{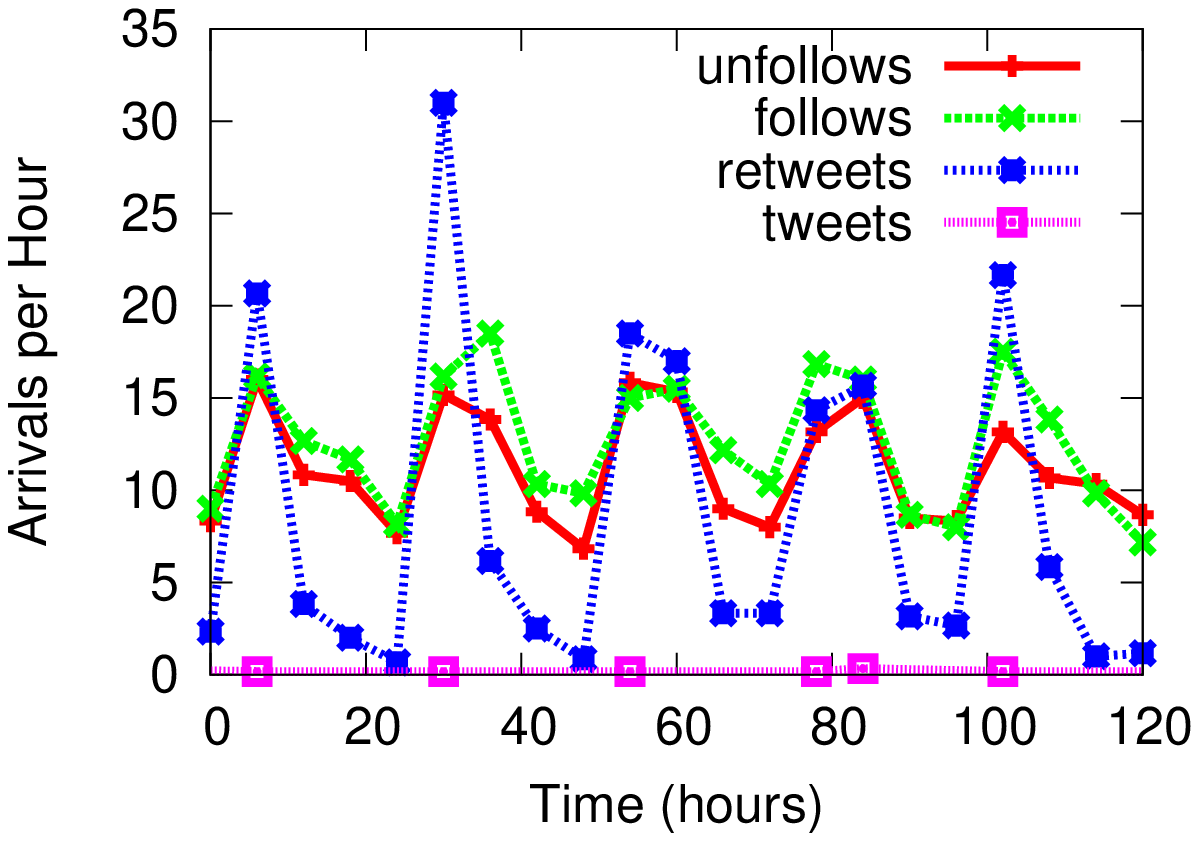}\label{subfig:u_2}}
    \subfigure[user with $d_{in}=112,988$]{\includegraphics[width=0.32\textwidth]{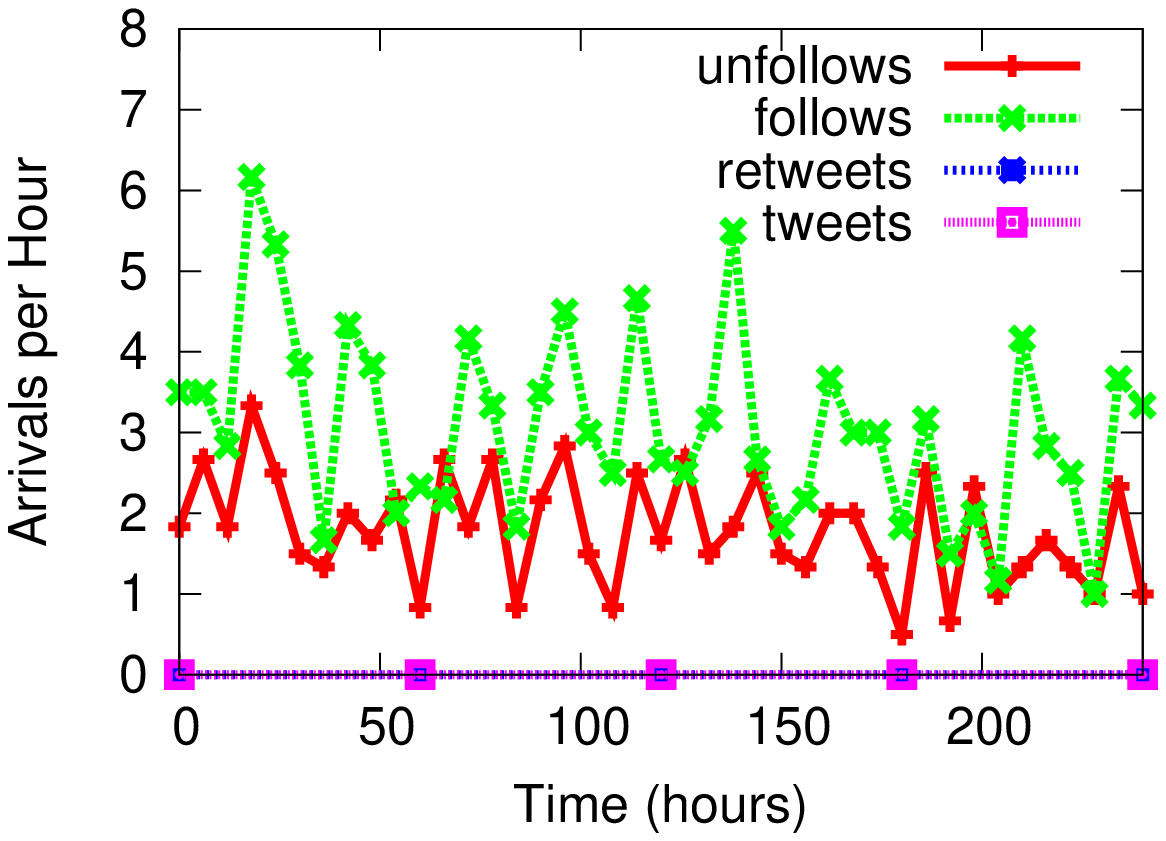}\label{subfig:u_3}}
    \caption{The arrival of follows, unfollows, tweets, and retweets as a function of time for several high-degree users ($d_{in}$ refers to the user's indegree).  Each plot represents a different individual user as she gains and loses followers, tweet, and is retweeted. (a) Around hour 110 the user receives a large number of retweets, which causes a large number of new followers. (b) A burst in retweets at hour 30 causes only a negligible increase in the new followers. (c) The user consistently receives new follows and unfollows without tweeting or being retweeted.}
    \label{fig:indiv_follow_arrival}
\end{figure*}

To gain intuition about temporal dynamics of Twitter user activity, in Figure \ref{fig:indiv_follow_arrival} we plot the arrivals of new follows and unfollows per hour for several high degree users as a function of time.  Also plotted against these arrival rates is how many times per hour the user is retweeted, as well as when they themselves tweet.  Each plot represents a different individual user as they gain and lose followers, tweet, and are retweeted over the hours of the month.  In all the plots we notice the presence of fluctuations with 24 hour periodicity.  These fluctuations correspond to the daily activity cycles on Twitter and represent the steady state of the Twitter network.

We also observe interesting deviations. For example, in Fig.~\ref{subfig:u_1}, around hour 110 the user receives a large number of retweets, which is later followed by a large number of new followers.  This is an example of an information diffusion event causing a perturbation in the new follower arrival rate.  However, retweets do not always lead to new follows. For example, in Figure~\ref{subfig:u_2} a burst in the number of retweets occurs around hour 30, but this causes only a negligible increase in the new follower arrival rate.  And to demonstrate that even users with no activity still gain and lose followers, Fig.~\ref{subfig:u_3} shows a user who consistently receives new follows and unfollows at a constant rate, even though she does nothing.

\xhdr{Detecting follow perturbations}  
Next we focus our analysis on two cases when steady follower arrival and departure rates change suddenly and abruptly. These abrupt changes are often the response to information diffusion. Our aim is to understand why some information diffusion events cause network changes while others do not. In order to do so, we first develop a method for identifying perturbations to the steady arrival of new follows and unfollows. To be more specific, we aim to identify periods of time in which a user receives more than a given threshold of follows/unfollows compared to what was expected historically.  

The biggest challenge associated with identifying these perturbations, or {\em bursts} as they will henceforth be called, is the periodic fluctuations in the arrival rate across the hours of the day.  To remove this periodicity, we initially employed traditional methods such as Fourier Transforms, but the abundance of noise in the arrivals as well the bursts themselves proved to be problematic.  Instead, we proceed as follows.

We treat the arrival of new follows and unfollows over the course of the month of each user as an independent time series: Let $\mathbf{x}=\left\{ x_1, x_2, ..., x_n\right\}$ be the number of new follows a user receives for each hour of the month.  (For the sake of clarity, we will describe our method using only follows, but the exact same analysis is applied to all users' unfollows independently.) We are interested in intervals of time in which the number of follows increases significantly more than expected, given the hour of day.  Let $t_i$ represent the $i^{th}$ hour of the month, and let $f(t_i)$ be the difference between actual new follows and expected follows during $t_i$:
\begin{align*}
f(t_i) &= x_i - E\left[x|h(t_i)\right]\\
 &= x_i - \frac{\sum_{j; |t_i - t_j| \leq 48,\, h(t_i) = h(t_j)} x_j \cdot w(t_i - t_j)}{\sum_{j; |t_i - t_j| \leq 48,\, h(t_i) = h(t_j)} w(t_i - t_j) }
\end{align*}

where $h(t)$ returns the hour of day in which time $t$ occurs, and $w(t)$ is an exponentially decaying weight function whose parameters are set using maximum likelihood.  Effectively, this is locally weighted regression, but only points at 24 hour periods are used to calculate the expected average.  

The function $f(t_i)$ now represents how many new followers a user received compared to the expected amount for hour $t_i$.  When $f(t_i)$ remains close to 0, this is considered to be the steady state behavior that most high degree users typically demonstrate.  However, we consider a {\em burst} to occur at time $t_i$ if $f(t_i)$ is greater than two standard deviations.

\hide{
We therefore choose a $\tau$ that maximizes the $\chi^2$ distribution for the co-occurrence of these pairs of bursts.  TODO:Find a better way to show Fig~\ref{ig:chisquare_bursts}
\begin{figure}
	\includegraphics[width=0.5\textwidth]{frac_non_null_chi_square_vs_thresh_per_user2}
	\caption{For each user and each threshold value that designates a burst from a normal fluctuation, the chi-square score between a retweet/tweet burst being followed by a follow/unfollow burst was performed.  Using a burst threshold of 4 standard deviations provides the maximum association between retweet and follow bursts TODO: Find a more intuitive version of this fig}
	\label{fig:chisquare_bursts}
\end{figure}
} 

It is worth noting that we also experimented with an alternative method for removing the periodicity in the follower dynamics. We employed the method proposed by Szabo and Huberman \cite{szabo2010predicting} where time is not counted in actual seconds but in the number of posts made on the entire site.  While the method removed some periodicity, we still found follow activity to be correlated with the hour of day.  We explain this by pointing out that most high degree users do have periodicity in the arrival of new followers, but this periodicity is out of sync with other users.  For example, a user in San Francisco who is followed by primarily local users would likely have a different hour of the day when they receive the most new follows compared to a user in England.  Therefore, for our dataset it is necessary to remove periodicity in a way that is independent for each user.

\xhdr{Burst co-occurrences} With a mechanism for detecting bursts, we now focus on instances in which bursts occur in close proximity to information diffusion events.  Changes in the arrival of new follows to a given user often coincide with large retweet cascades generated by a tweet posted by the user.  We apply the normalization process to a user's arrival of new follows, unfollows, tweets, and retweets, all independently. We call {\em retweet-follow burst} any time a (2 standard deviation) burst in retweets occurs in one hour, and then a burst in follows occurs within the next hour. Additionally, tweeting too much (or more than the user normally does) can annoy user's followers and thus cause unfollows.  Therefore, we also focus on {\em tweet-unfollow bursts}: a burst in tweets occurs within an hour of a burst in unfollows.  As we shall show next, these two types of bursts are not only common, but they can explain many changes in the user's local network.

\begin{figure}[!t]
    \centering
    \begin{tabular}{ c c}
    \textbf{ Tweet-Unfollow Bursts} & \textbf{ Retweet-Follow Bursts} \\
    \hline
    \subfigure[]{\includegraphics[width=0.23\textwidth]{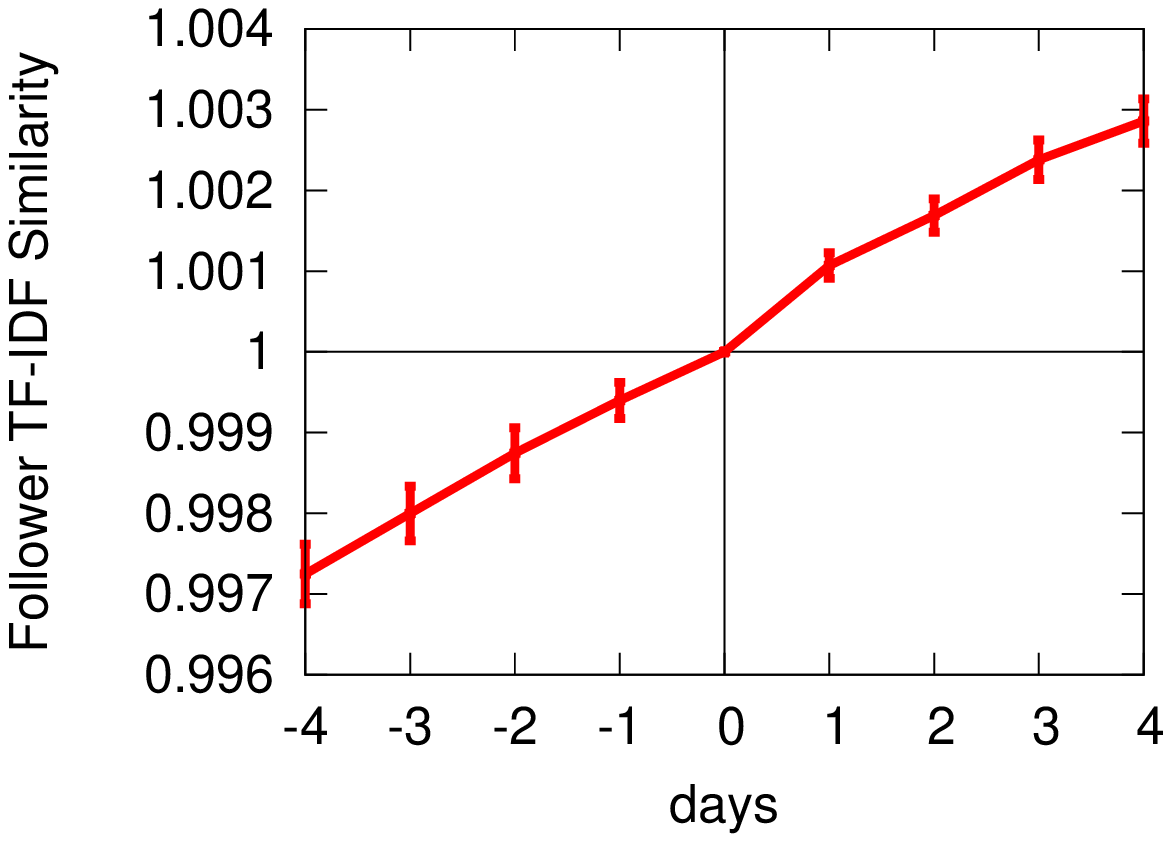}\label{subfig:tfidf_around_tw_spike}} &
    \subfigure[]{\includegraphics[width=0.23\textwidth]{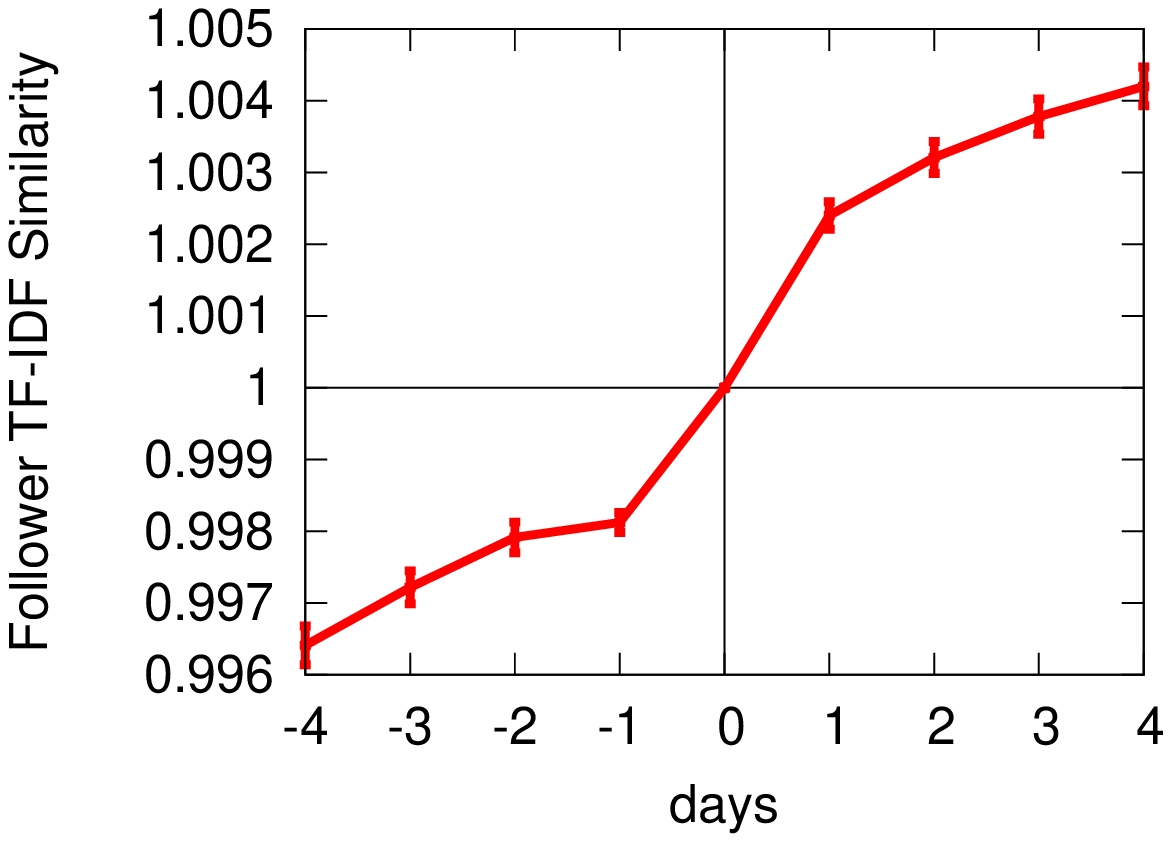}\label{subfig:tfidf_around_rt_spike} }\\
    \end{tabular}
    Follower Tweet Similarity
    \begin{tabular}{ c c}
    \subfigure[]{\includegraphics[width=0.23\textwidth]{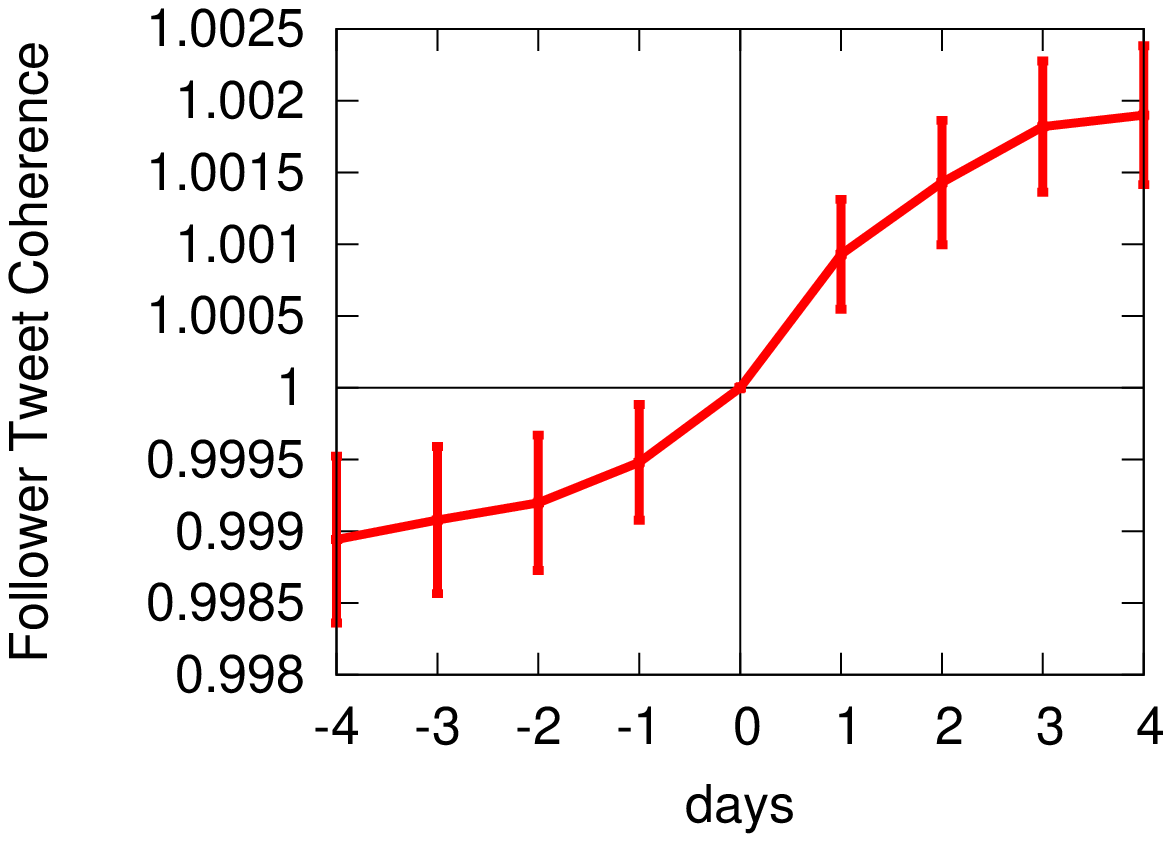}\label{subfig:tfidf_variance_around_tw_spike}} &
    \subfigure[]{\includegraphics[width=0.23\textwidth]{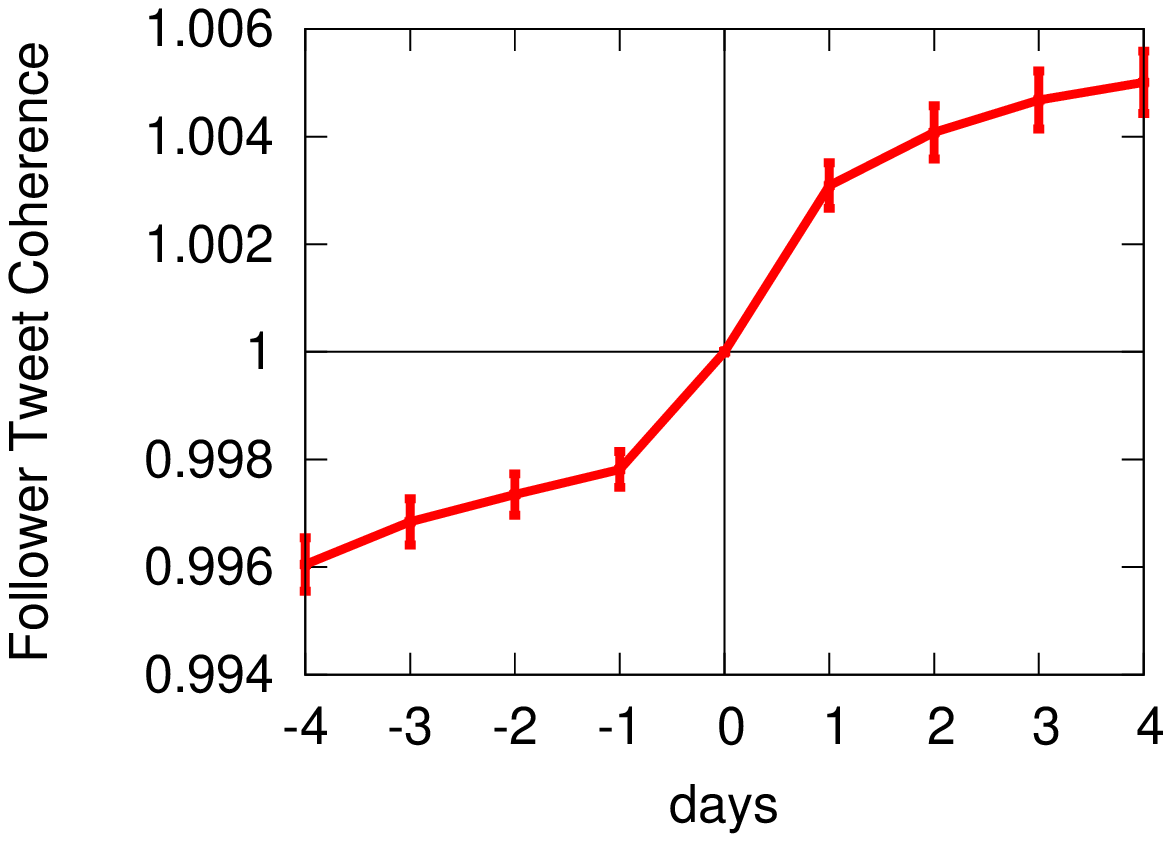}\label{subfig:tfidf_variance_around_rt_spike} }\\
    \end{tabular}
    Follower Tweet Coherence
    \begin{tabular}{ c c}
    \subfigure[]{\includegraphics[width=0.23\textwidth]{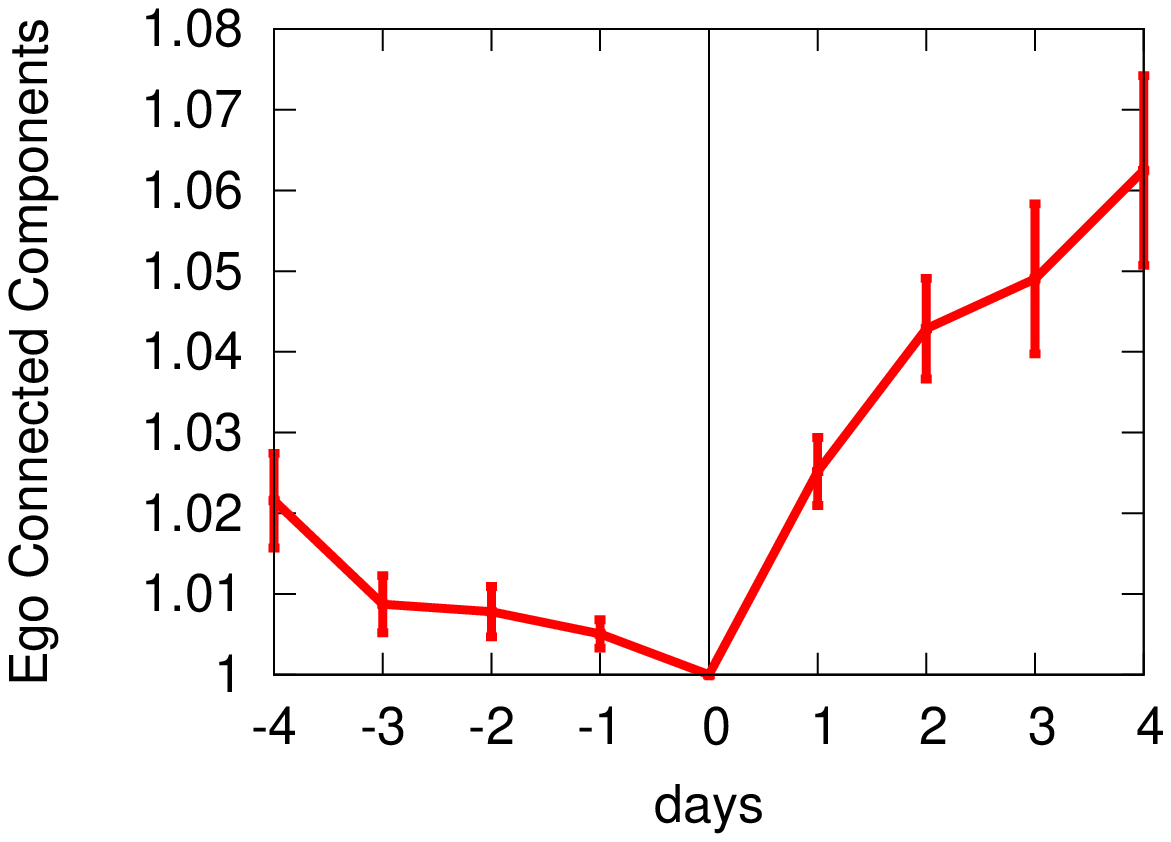}\label{subfig:egoWCC_around_tw_spike}} &
    \subfigure[]{\includegraphics[width=0.23\textwidth]{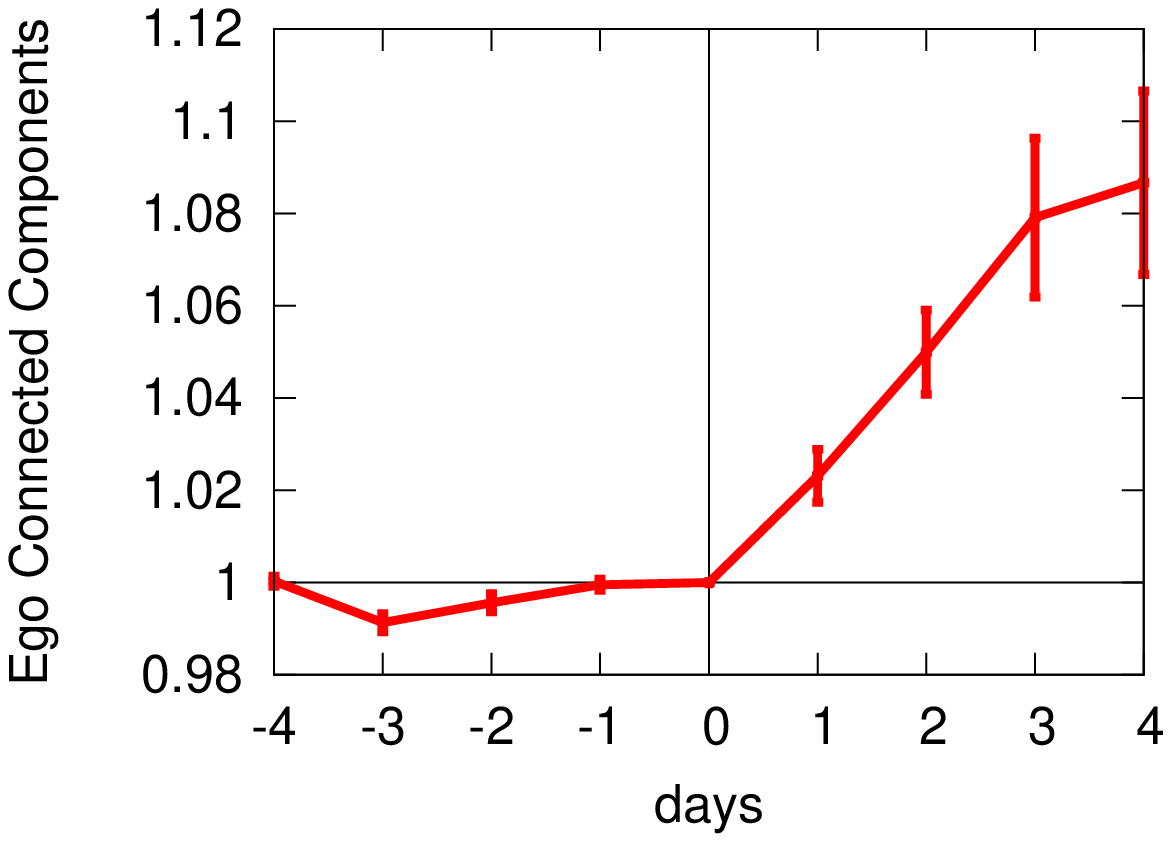}\label{subfig:egoWCC_around_rt_spike}} \\
    \end{tabular}
    Follower Weakly Connected Components
    \begin{tabular}{ c c}
    \subfigure[]{\includegraphics[width=0.23\textwidth]{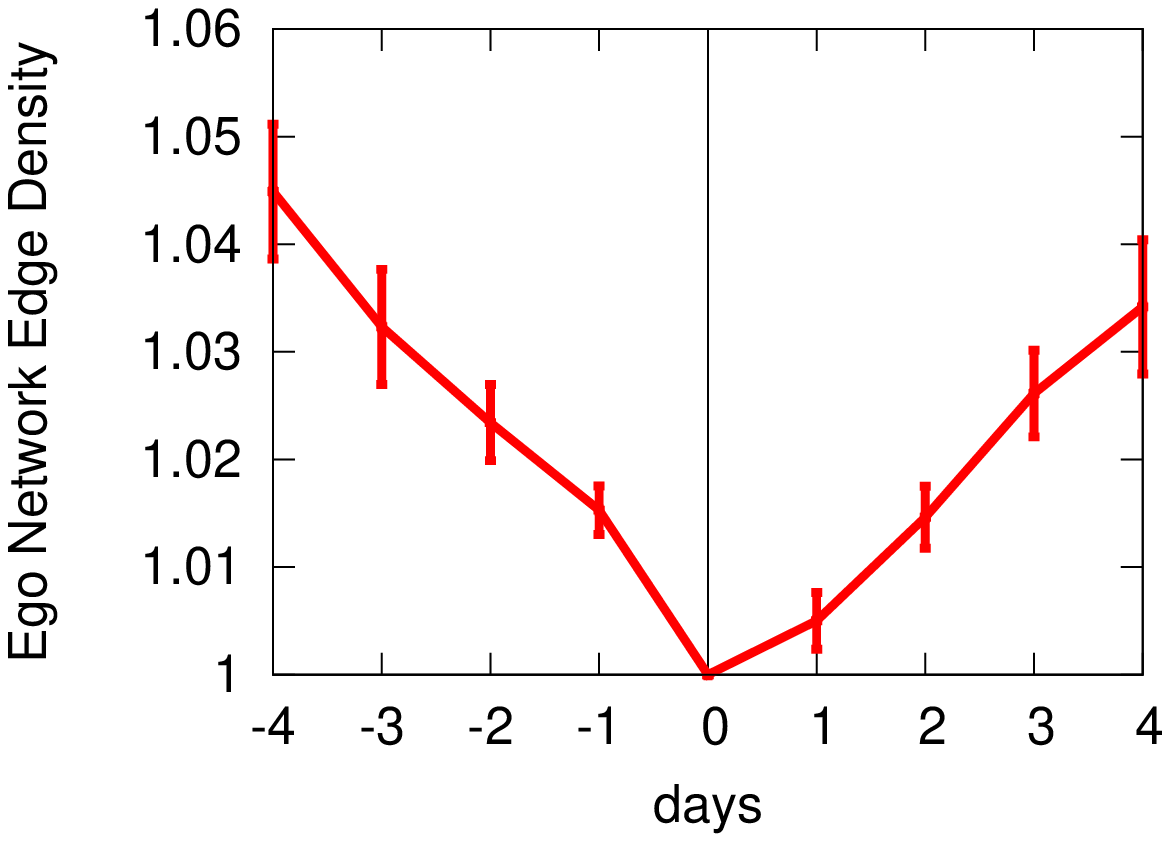}\label{subfig:egodensity_around_tw_spike}} &
    \subfigure[]{\includegraphics[width=0.23\textwidth]{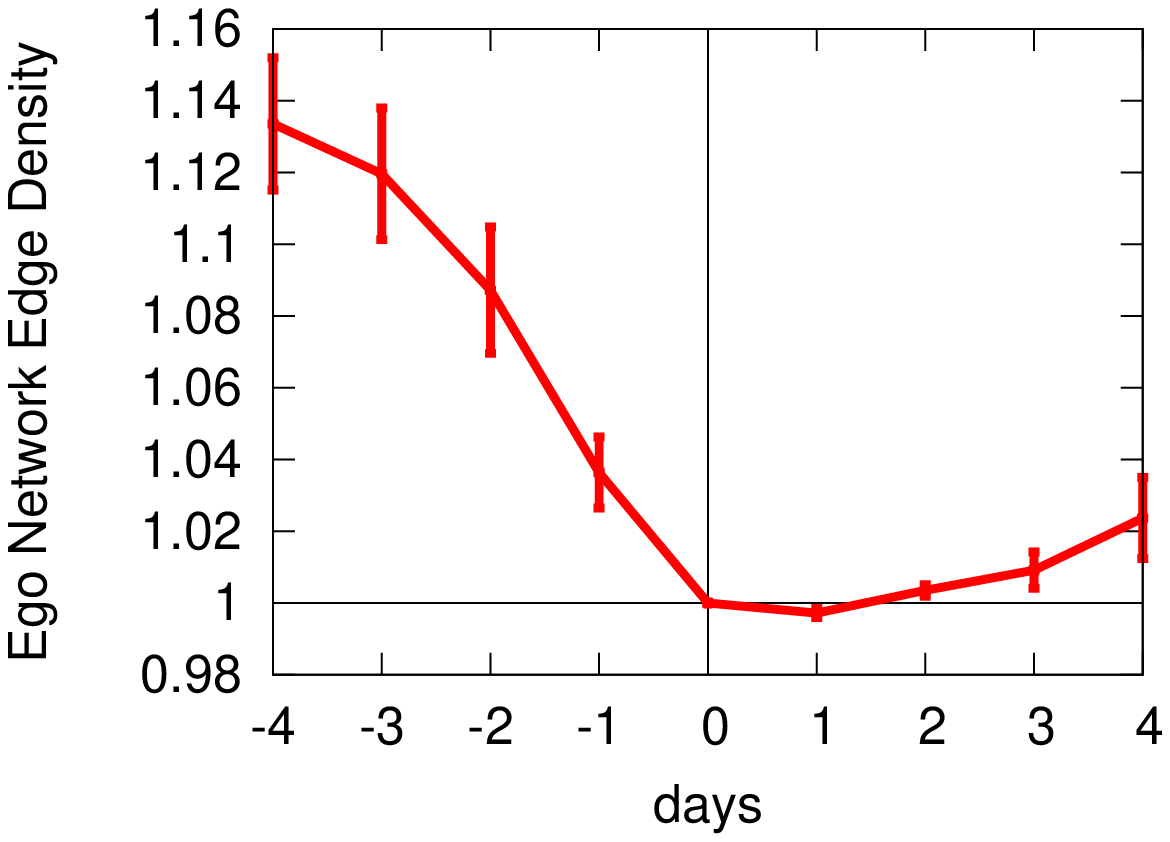}\label{subfig:egodensity_around_rt_spike}} \\
    \end{tabular}
    Follower Edge Density
    \caption{Metrics of user's follower ego-network around the time of a retweet-follow and a tweet-unfollow burst. For 4 days preceding and succeeding a burst, we plot the average value relative to the metric at the moment of the burst.  For example, if at day = -2, the metric is 1.02, that means on average 48 hours before the burst the metric was 2\% greater than what is was on the day of the burst.  Overall, bursts tend to increase the coherence of the local network in both the interests of the connected users as well as the number of mutual connections between them.}
    \label{fig:tfidf_spikes}
\end{figure}

\subsection{User's ego-network during a burst}

Through the process described above, we examine all users with at least 2000 followers and identify 2.1 million instances where a retweet burst was immediately followed by a follow burst.  We focus on high degree users because for low degree users, the arrival in new follows or unfollows is not frequent enough, so detecting sudden changes is not reliable. Our analysis focuses on the ego-network of a user: the subgraph composed of a user's followers (excluding the user herself) and all the follower relationships between them. Through examining the properties of users' ego-networks before and after the occurrence of the bursts, we show that bursts contribute to network evolution by advancing it in abrupt intervals.

\xhdr{The similarity of follower tweets}  The first question we ask is how similar a user is to her followers, and how this similarity changes during a burst.  More specifically, for a pair of users we want to quantify how similar are their interests in different types of information.  To do this, we measure the  textual similarity of their tweets. The more similar the tweets of users, the more similar the information that diffuses through them. For each user, we aggregate every tweet she posted during the month into a single ``document.''  We define the {\em user tweet similarity} as the cosine similarity of the TF-IDF weighted word vectors between the two users' aggregated tweet documents. Although simple, this method provides a robust measure of similarity between a pair of users.  We note that the aggregated tweet documents also contain retweets, but retweets only account for a small fraction of all tweets.  Thus, a user's tweet document is largely unaffected by any retweets they might have made.

Using the tweet similarity of a user and their followers before and after a burst occurs, we investigate whether users' followers become more or less similar in the content of their tweets.  
For each burst, we measured the average follower similarity at 24 hour intervals for multiple days before and after the burst.  We need to average this metric across all users and bursts, so we normalize each measurement by its value exactly at the time of the burst.  Thus, the metric is comparable across all users.  Figures~\ref{subfig:tfidf_around_tw_spike}, \ref{subfig:tfidf_around_rt_spike} show the results averaged across all bursts of the respective types.  If the $y$-axis is 1, then the metric for that period of time was the same value at the time of the burst.  For both types of bursts, we observe a statistically significant increase in the user similarity, but for the retweet-follow bursts there is an abrupt increase.

Overall, we find that during a retweet-follow burst, the follower tweet similarity increases abruptly, and to a lesser extent, for tweet-unfollow bursts.  On average the follower tweet similarity increases over the course of the month, which implies that most users' followers become more congruent to them.  However, this rate of increase speeds up by 25.5\% during a retweet-follower burst, shown in Table~\ref{tab:during_burst}. 

The reason for this acceleration of change is the nature of new followers gained during a burst versus the new followers that are not gained through information diffusion.  New followers that a user gains through being retweeted have a 76.6\% higher tweet similarity than new followers that never were exposed to a retweet.  Additionally, the new followers gained through retweets are 109.5\% more similar than pre-existing followers.  This means that, during a retweet-follow burst, a user gains followers more similar to her than she would normally gain, and these followers are even more similar than the followers she already has.


On the other hand, for the case of tweet-unfollow bursts, a user's followers tweet similarity increase rate speeds up by only 11\% (see Table~\ref{tab:during_burst}). This increase, however, is caused by current followers with less similar tweets, who then unfollow the user.  Indeed, the tweet similarity of a user who unfollows is 36.1\% lower compared to a follower that endures the entire month.
 
To rule out spurious causes of the increase in tweet similarity, we conducted a separate randomized experiment where we ran the above analysis again, but this time for each actual follow/unfollow, we randomized the recipient of the action but preserved the source user.  In other words, if the data contains the event ``user $A$ follows user $B$'', we would replace user $B$ with another randomly selected user in the network.  Here, the tweet similarity between users and their followers {\em decreased} significantly, which means that the increased similarity observed above is not spurious.

We conclude that during the retweet-follow and tweet-unfollow bursts, the similarity of interests of a user's followers increases, and so the user's network becomes more homogeneous. This means that bursts cause a sudden ``jump'' in the network's evolution toward bringing similar users together and pushing dissimilar users farther apart.

\xhdr{Follower tweet coherence} To test the idea that followers become more related to {\em one another} (not just to the user) during a burst, we measure the similarity in tweets between the followers of a user.  Using the same method of TF-IDF cosine similarity of tweet content, we measured the similarity across all pairs of followers of a given user in the days succeeding and preceding the burst.  These measurements were normalized by their value during the burst, just as they were in the previous section.  We refer to this as {\em follower tweet coherence} and plot the results in Figures~\ref{subfig:tfidf_variance_around_tw_spike} and~\ref{subfig:tfidf_variance_around_rt_spike}. We observe that before the burst, the follower coherence steadily increases.  But just as the follower tweet similarity suddenly increases during the burst, so do the follower tweet coherence.  Both types of bursts cause the followers' tweets to become more aligned with each other. 

\xhdr{Connected components amongst followers}
We also examine structural changes to a user's local network neighborhood before and after a burst.
In particular, we study the number of weakly connected components (WCC) of the ego-network during a burst.  If the number of connected components is high, then the subgraph of followers is fragmented.  This would indicate that user's followers do not belong to a single cohesive community and tend not to follow each other.

We discover that bursts cause a large increase in the number of connected components in the network.  Relative to a user's ego-network WCC change over the entire month, a retweet-follow burst causes the arrival of WCC's to increase by 17.4 times during the burst, while tweet-unfollow burst increases the arrival by 4.0 times (see Table~\ref{tab:during_burst}). 
Furthermore, Figures~\ref{subfig:egoWCC_around_tw_spike} and \ref{subfig:egoWCC_around_rt_spike}
show the relative number of WCC's in the days proceeding and succeeding bursts.  In both cases, the number of connected components in the follower ego-network increases for several days after the burst.  Thus, we conclude that bursts cause an influx of new followers from unaffiliated communities into a user's follower ego-network.

\xhdr{Followers following each other}
Last, we analyze the follower ego-network edge density.  For a given set of followers, the metric represents what fraction of potential following relationships actually exist.  If this value is low, then a user's followers tend not to follow each other, whereas a high value would indicate a well connected ego-network.  In general, the ego-edge density for followers decreases, largely because users' ego-networks grow in the number of nodes over time.

We find that during both retweet-follow bursts and tweet-unfollow bursts, the density decreases significantly faster than when no bursts occur. As shown in Table~\ref{tab:during_burst} the density decreases 41.1 times faster for retweet-follow and 14.6 times faster for tweet-unfollow burst when compared to the period of no bursts.

Even more interesting is the change in ego-network density in the days around the burst. Figures~\ref{subfig:egodensity_around_tw_spike} and \ref{subfig:egodensity_around_rt_spike} plot the relative change in density for days before and after a burst.  Before either type of a burst, there is a steady decrease in density.  For the days after the burst, however, we observe interesting behavior.  For the retweet-follower burst, the density actually remains nearly constant, while for the tweet-unfollow burst, the density actually increases, effectively as fast as it decreased before the burst.  
We explain the two observations as follows: during either type of a burst, there is a large change in the population of users in the ego-network, which results in a decrease in edge density.  In some sense, the burst is a shock to the structure of the ego-network.  In the days after the burst, however, the ego-network begins to densify as followers begin to connect to each other.  
After a burst, tweet similarity of the ego-network increases (Figures~\ref{subfig:tfidf_around_tw_spike}, \ref{subfig:tfidf_around_rt_spike}).  As this happens, the followers themselves become more related to each other.  With highly related users close to each other in the network, they are more likely to discover and follow each other, thereby increasing the density of the ego-network.

\begin{table}[t]
\begin{tabular}{|c|c|c|}
\hline
\textbf{Metric} & \begin{minipage}[b] {.15\textwidth}\centering \textbf{ $\Delta$ During Burst Retweet-Follow Burst} \end{minipage}& \begin{minipage}[b]{.15\textwidth}\centering \textbf{ $\Delta$ During Spike Tweet-Unfollow Burst} \end{minipage}\\
\hline
Tweet similarity & 25.5\%& 11.8\% \\
Components & 1737\% & 399.8\% \\
Edge density & -4199\% & -1467 \%\\
\hline
\end{tabular}
\caption{ 
How much does the rate of change of each metric accelerate/decelerate during a burst? Tweet similarity slowly increases over time, but this increase is accelerated by 25.5\% on average when a user is undergoing a retweet-follow burst. The number of connected components also sharply increases.
Edge Density is slowly {\em decreasing}, but during a retweet-follow burst this rate of decrease is 4,199\% faster.}
\label{tab:during_burst}
\end{table}

\subsection{Tweet content and bursts}
\label{sec:textanalysis}
Now that we have a better understanding of the network effects of a bursts, we address the content of the information causing the bursts.  We ask the question whether there are certain types of content that are more likely to cause a burst in new followers.

%
To study this question, we iterated across all instances of retweet bursts and extracted the text of the tweet creating the burst.  For each token that occurred in these tweets, we measured whether or not the presence of the token increased or decreased the probability that the tweet would cause a burst in new followers.  We filtered out tokens which were present in less than 10 tweets.  We then identified all tokens that violated the null hypothesis of having no effect on new follow burst probability, using the Pearson $\chi^2$ test with 95\% confidence.  
Note that all these tweets caused a burst of retweets, but only a fraction of them lead to a burst in new followers.  These tokens thus had a statistically significant effect on whether or not the tweet would cause a burst in new followers.  We then rank these tokens by the ratio 
\begin{align*}
R(tok_i) = \frac{ Pr( \mbox{new follower burst occurs} \, | \, tok_i\mbox{ in tweet}  ) }{ Pr( \mbox{new follower burst occurs} \, | \, tok_i \mbox{ not in tweet }  ) }
\end{align*}
for all tokens $tok_i$.  This ratio quantifies how much the presence of a particular token within a tweet will increase the chances of a follower burst.




\begin{table}[t]
\centering
\begin{tabular}{|c | c |}
\hline
\textbf{Token} & \textbf{Prob. Ratio}\\
\hline
officer & 2.9082 \\
officers & 2.5901 \\
\#n30 & 2.5655 \\
\#occupyphilly & 2.5599 \\
LAPD & 2.4942 \\
solidarity & 2.4847 \\
eviction & 2.4675 \\
riot & 2.2935 \\
protestors & 2.1301 \\
@occupyla & 2.1290 \\
police & 2.0845 \\
\#nov30 & 2.0406 \\
arrest &  2.0179 \\
cops & 1.9983 \\
\#ows & 1.5879 \\
protesters & 1.5278\\
\hline
\end{tabular}
\caption{
Many of the top 100 tokens that had the most relative increase in the probability of causing a follower burst were associated with the Occupy Wall Street movement.
}
\label{tab:ows_tokens}
\end{table}

All tweets in our dataset were created in November of 2011, at which time the ``Occupy Wall Street'' movement was less than two months old.  Occupy Wall Street was a protest movement against income inequality, and on several occasions the protests led to conflicts with police.  Of the top 100 tokens with the largest probability increase ratio, at least 16 of them are associated with this movement. These 16 tokens are listed in Table \ref{tab:ows_tokens}.  Additionally, a tweet containing the top token ``officer'' is almost {\em three times} more likely to cause a new follow burst than if the tweet did not contain the token.

On the other hand, we find that there are two types of events that generally cause large unfollow bursts. As expected, the first is when tweets include content that is either offensive (such as obscenities) or is spam.  For example, tokens such as ``free'', ``sale'', and ``download'' all increase the probability of an unfollow burst. Interestingly, the second is when sports stars change teams. In our data we observed several instances when a professional sportsman would announce switching his team, which in turn would lead to a huge shift in his follower base: supporters of his old team would unfollow, while supporters of the new team would create new follower links.

\section{Modeling Follower Bursts}
\label{sec:model}
So far, we have seen how information diffusion events such as large retweet cascades can cause sudden bursts in network dynamics.  To better understand the effect of bursts, we develop a model that not only explains these observations but can predict them as well.  Therefore, our goal is to develop a model that, given a set of retweet bursts, predicts which bursts will cause a spike of new followers.  The model we present here does not consider the content of the tweets but still performs well in practice.  The process that inspires our model is not applicable to bursts in tweets leading to bursts in unfollows.  

Most new follows are the result of a triadic closure: if  user $i$ creates a new follow edge to another user $k$, then often there exists at least one ``intermediate'' follower $j$ (as illustrated in Figure~\ref{fig:rt_follow_example}(right)).  In fact, across all new follows in our dataset, the average directed shortest path between the users just before the creation of the new follow edge is $2.036\pm0.007$. (If all new follows would be a result of a triadic closure then shortest path before the follow edge creation would be 2.)  In this light, we can safely assume that all of a user's new follows come from users who already follow one of her followers (but do not directly follower her already).  We  refer to this set as a user's {\em 2-hop neighborhood}, and we call the user's current set of followers the {\em 1-hop neighborhood}. 

Now, in order to predict the occurrence of follow bursts, we need to model the probability of a user transitioning from the 2-hop neighborhood to the one 1-hop neighborhood during a large information diffusion event.

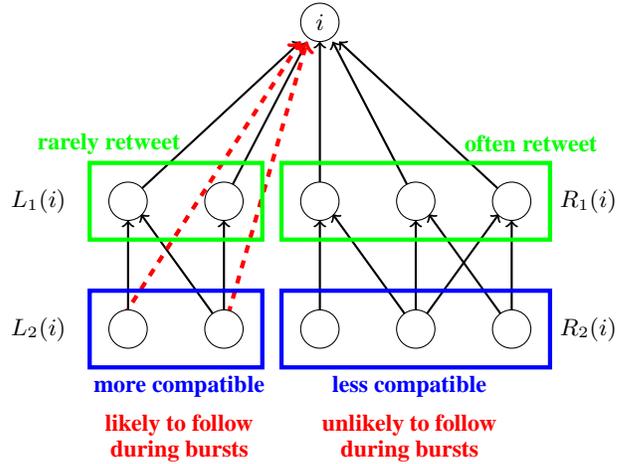
\begin{figure}[t]
\centering
\begin{tikzpicture}[scale=.85]

\draw [->, thick] (-3, -2)--(-.3,.5) ;
\draw [->, thick] (-1.5, -2)--(-.2,.5) ;
\draw [->, thick] (0, -2)--(0,.5) ;
\draw [->, thick] (1.5, -2)--(.2,.5) ;
\draw [->, thick] (3, -2)--(.3,.5) ;

\draw [->, thick] (-3, -4)--(-3,-2.3) ;
\draw [->, thick] (-1.5, -4)--(-1.5,-2.3) ;
\draw [->, thick] (-1.5, -4)--(-2.8,-2.2) ;
\draw [->, thick] (0, -4)--(0,-2.3) ;
\draw [->, thick] (1.5, -4)--(.2,-2.2) ;
\draw [->, thick] (1.5, -4)--(1.5,-2.3) ;
\draw [->, thick] (1.5, -4)--(2.8,-2.2) ;
\draw [->, thick] (3, -4)--(3,-2.3) ;
\draw [->, thick] (3, -4)--(1.7,-2.2) ;

\draw [->, dashed, red, ultra thick] (-3.2, -4)--(-.3,.5);
\draw [->, dashed, red, ultra thick] (-1.5, -4)--(-.2,.5);

\draw [green, ultra thick] (-3.6, -1.4) rectangle (-.9, -2.6);
\draw [green, ultra thick] (-.6, -1.4) rectangle (3.6, -2.6);

\draw [blue, ultra thick] (-3.6, -3.4) rectangle (-.9, -4.6);
\draw [blue, ultra thick] (-.6, -3.4) rectangle (3.6, -4.6);

\draw[fill=white] (-3,-2)  circle (.3);
\draw[fill=white]  (-1.5,-2) circle (.3);
\draw[fill=white]  (0,-2) circle (.3);
\draw[fill=white]  (1.5,-2) circle (.3);
\draw[fill=white]  (3,-2) circle (.3);

\draw[fill=white] (-3,-4) circle (.3);
\draw[fill=white]  (-1.5,-4) circle (.3);
\draw[fill=white]  (0,-4) circle (.3);
\draw[fill=white]  (1.5,-4) circle (.3);
\draw[fill=white]  (3,-4) circle (.3);

\draw [green, fill=white](-3.3, -1.1)node(){\textbf{rarely retweet}};

\draw [green, fill=white](3.3, -1.1)node(){\textbf{often retweet}};

\draw [blue, fill=white](-2.2, -4.9)node(){\textbf{more compatible}};

\draw [blue, fill=white](1.4, -4.9)node(){\textbf{less compatible}};

\draw [red, fill=white](-2.2, -5.5)node(){\textbf{likely to follow} };
\draw [red, fill=white](-2.2, -5.9)node(){\textbf{during bursts} };

\draw [red, fill=white](1.4, -5.5)node(){\textbf{unlikely to follow} };
\draw [red, fill=white](1.4, -5.9)node(){\textbf{during bursts} };

\draw[](-4.4,-2)node(){$L_1(i)$};
\draw[](-4.4,-4)node(){$L_2(i)$};

\draw[](4.2,-2)node(){$R_1(i)$};
\draw[](4.2,-4)node(){$R_2(i)$};

\draw[fill=white] (0,.8) node(){$i$}circle (.3);

\end{tikzpicture}
\caption{  An example of a user $i$ who would have a high probability of experiencing a burst.  There is a subset of users in $N_2(i)$ who have a high tweet similarity with $i$ and thus would be more compatible followers if they were exposed to her tweets, but the users in $N_1(i)$ that they follow do not retweet so this exposure has not happened.  The burst would occur when this subset finally sees a retweet during a retweet burst and they subsequently follow $i$.  On the other hand, the other users in $N_2(i)$ have most likely already seen $i$'s tweets, and so they are less likely to form a new follow during a burst.  }
\label{fig:model_explain}
\end{figure}

A retweet follow burst is defined as a sudden increase in new follows to a user.  These new follows would not normally occur during steady-state behavior, but instead represent a set of potential-followers discovering the user for the first time, thanks to an information diffusion event.  Therefore, the key to predicting a retweet-follow burst is identifying a set of similar users who would not normally be exposed to the user but respond to the burst in retweets.  Figure~\ref{fig:model_explain} illustrates the process. 

Consider the case when user $i$ is experiencing a retweet-burst by her tweet being propagated through the network. Then there is a set $L_2(i)$ of 2-hop neigbhors of $i$ that have very compatible interests with $i$.  However, users in $L_2(i)$ only have access to $i$ through a set of 1-hop neighbors $L_1(i)$ who rarely retweet $i$. Thus, users in $L_2(i)$ are unlikely to know about $i$ as they have not been exposed to $i$. Users in $L_2(i)$ are unique to other 2-hop neighbors (e.g., $R_2(i)$) that either have incompatible interests with $i$ or have already been exposed to $i$'s tweets but have chosen not to follow her.  

A burst of new follows then happens when users in $L_2(i)$ are exposed to $i$'s content for the first time (via a retweet cascade), and as a result they are likely to follow user $i$.  

Next, we use this intuition to develop the follower burst prediction model, but first we must quantify what it means for two users to have compatible interests.

\xhdr{Tweet similarity drives probability of new follow}  As discussed in the previous section, new follows tend to increase the average tweet similarity between a user and her followers, and this is especially the case during a retweet-follow burst.  We take the tweet similarity between two users as a simple but effective measure of how compatible their interests are, and thus how compatible they would be as followers of each other. 
Thus, we expect the probability of user $i$ following user $k$ to be high, if the newly created edge $i \rightarrow k$ would increase the overall tweet similarity of $k$'s followers.

\begin{figure}
\includegraphics[width=.5\textwidth]{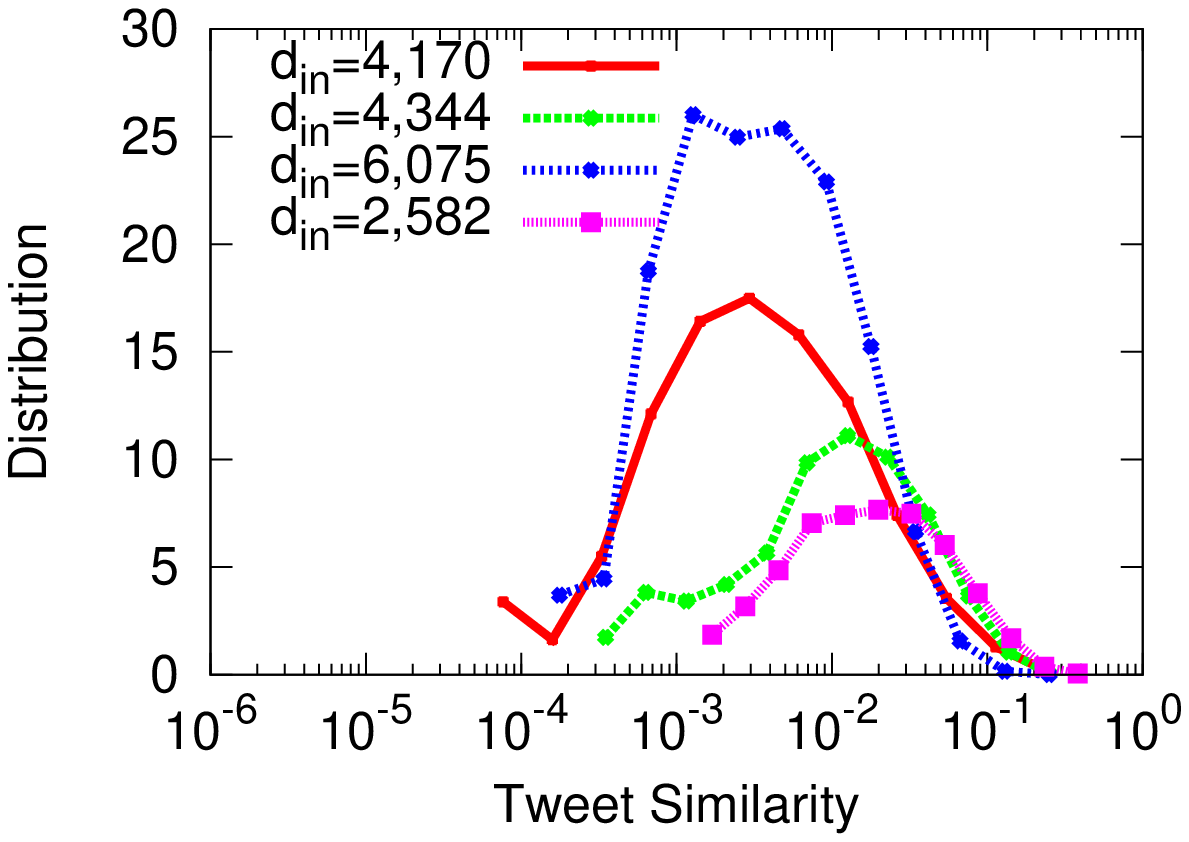}\label{subfig:tweet_sim_dist}
\caption{The distribution of follower tweet similarity for several different users.  The number of followers for each user is listed in the legend.  Even for users with comparable number of followers, the variability in the distribution of follower tweet similarity is significant.}
\label{fig:tweet_sim_dist}
\end{figure}

Increasing the average follower tweet similarity does not have the same effect on all the users.
For example, Twitter user @cnnbrk (CNN Breaking News) might be followed by a wide range of users who simply want to stay informed about news events. On the other hand followers of @espn (ESPN News) will likely have a more narrow interest in sports.  
Therefore, a new follower of @cnnbrk has to have a lower absolute tweet similarity in order to increase the average tweet similarity of @cnnbrk's followers. On the other hand, a new follower would have to be very interested in sports in order to increase the average tweet similarity of @espn's followers.  

Figure~\ref{fig:tweet_sim_dist} confirms this intuition by plotting the distribution of tweet similarities of followers for several different users.  Notice a high variability in the distribution of follower tweets for individual users. In order to reliably model the probability of a new follow we now account for this variability.

\xhdr{Comparing similarity across all users} We take advantage of the fact that the distribution of follower tweet similarity follows a log-normal distribution for each user.  As illustrated in Figure~\ref{fig:tweet_sim_dist} the tweet similarity $S(i,j)$ between user $i$ and her follower $j$ follows a log-normal distribution: $\ln\left[ S(i,j) \right] \sim \mathcal{N}\left( \mu_i, \sigma_i^2\right)$. And so it follows for all users $i$
\begin{align*}
Y{ij}\equiv &\frac{ \ln\left[S(i,j)\right] - \mu_i }{\sigma_i} \sim \mathcal{N}(0,1) \\
\mbox{with}& \\
\mu_i =& \frac{1}{|N_1(i)|}\sum_{k\in N_1(i)} \ln\left[S(i,j) \right] \\
\sigma_i^2 =&\frac{1}{|N_1(i)|}\sum_{k\in N_1(i)} \left(\ln\left[S(i,k) \right] - \mu_i\right)^2.
\end{align*}
where the set $N_1(i)$ is the current 1-hop neighborhood of user $i$.  Now, the value $Y_{ij}$ quantifies where in the distribution of follower tweet similarity user $j$ would be if she chose to follow user $i$. 
If $Y_{ij}=0$, then $j$'s similarity to $i$ is equal to the average similarity of $i$'s followers.
However, the sign as well as the magnitude of $Y_{ij}$ quantify how much more/less similar $j$ is when compared to an ``average'' follower of $i$. This way $Y_{ij}$ is normalized and comparable across all users.

\begin{figure}
    \centering
    \includegraphics[width=0.5\textwidth]{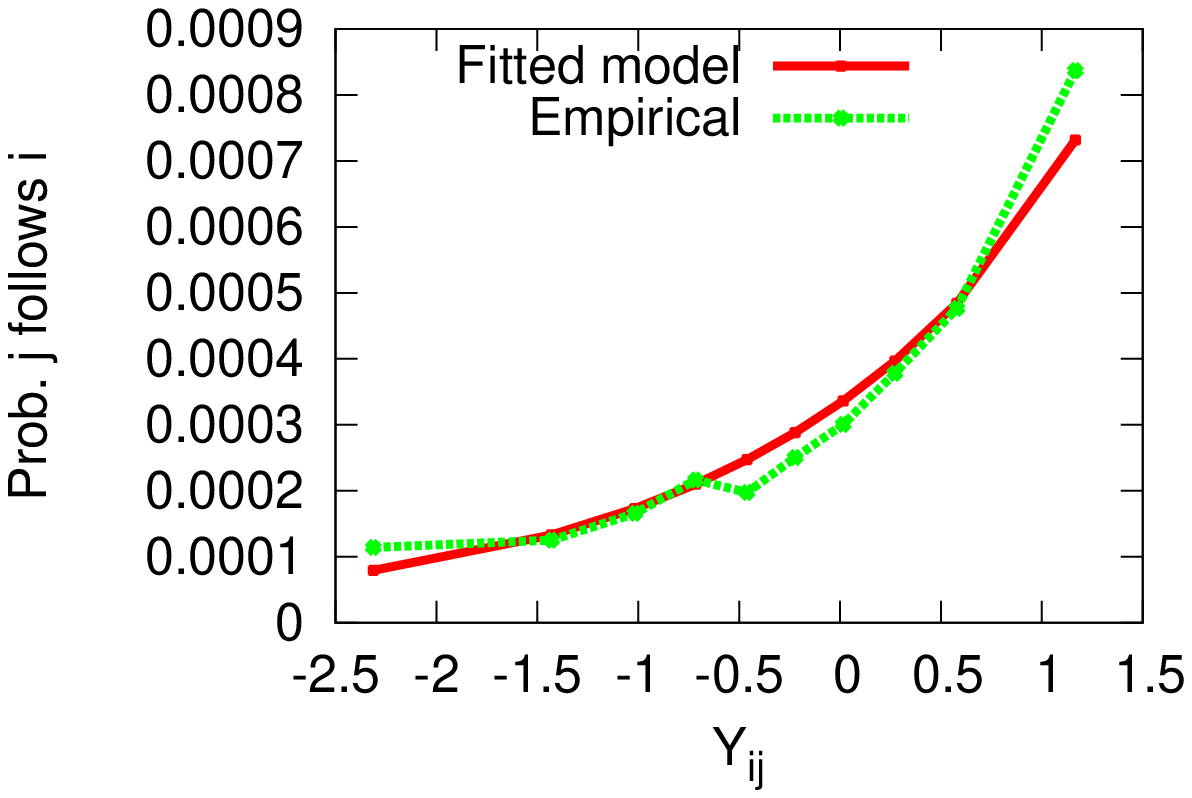}
\caption{The normalized log-tweet similarity between two users $i$ and $j$ plotted against the probability that $j$ will follow $i$, given that $i$ recently tweeted. }
    \label{fig:hop2to1}
\end{figure}

Now, finding the probability of a new follow as a function of $Y_{ij}$ is a matter of empirical observation.  Figure~\ref{fig:hop2to1} plots $Y_{ij}$ against the average probability of $j$ following $i$ within 3 days of $i$ tweeting. 
We observe (and a likelihood test confirms) that $P(j\mbox{ follows }i \,|\, Y_{ij})$ is an exponential function of $Y_{ij}$.  Plotted along with the empirical observation of $P(j\rightarrow i \,|\, Y_{ij})$ is the fitted curve:
\begin{align}\label{eqn:pij}
\hat{P}_{j,i}\equiv&\,C\cdot \exp\left( \alpha \cdot Y_{ij} \right) \nonumber\\
=&\,C \cdot \left[ \frac{S(i,j)}{\exp(\mu_i)} \right]^{\alpha / \sigma_i}
\end{align}
where $C=3.32\times10^{-4}$ and $\alpha = 0.6445$ were set using maximum likelihood.

The above result says that the probability of a user moving from the 2-hop neighborhood to the 1-hop neighborhood (\ie, forming a new follow edge) increases {\em exponentially} with normalized tweet similarity $Y_{ij}$.
Moreover, the result also explains why a user's average follower tweet similarity tends to increase over time: A potential new follower $j$, who is more similar to $i$ than $i$'s current followers, is almost an order of magnitude more likely to follow $i$ than someone who is dissimilar.

\xhdr{Predicting bursts}  With $\hat{P}_{j,i}$ given by Equation~\eqref{eqn:pij}, a possible way to predict the total of number of new follows a user receives during an information diffusion event would be to compute the expected number of new follows:
\begin{align*}
\sum_{j\in N_2(i)}\hat{P}_{j,i}.
\end{align*}
where $N_2(i)$ is the 2-hop neighborhood of user $i$.  This quantity, however, is a poor predictor of follow bursts.  In fact, the larger it is, the {\em less} likely a given retweet burst will produce a new follow burst.  The reason is that $\hat{P}_{j,i}$ is the probability of a new follow that is not conditioned on the occurrence of a retweet burst.  It quantifies the arrival of new follows during a steady-state, non-burst behavior.  If, for example, $\hat{P}_{j,i}$ is high and a retweet burst occurs, there is a high chance that user $j$ has already made the decision whether or not to follow $i$.  A retweet-follow burst, on the other hand, is the {\em interruption} of steady-state network dynamics.  A retweet-follow burst occurs when users in $N_2(i)$ with high tweet compatibility to $i$ are exposed to $i$'s tweet for the first time.  

A retweet-follow burst occurs when a sudden burst in retweets reaches a set of potential followers that are more compatible with the user than the typical users who are usually reached.  The more compatible the set of users reached during the retweet, and the less compatible the set of users that are normally reached, the more likely a burst is to occur.  Let $N_{RT}(i,[t, t+\Delta t))$ be the set of users in the 2-hop neighborhood who follow someone that retweeted user $i$ during the time interval $[t, t+\Delta t)$. In other words, users in $N_{RT}(i,[t, t+\Delta t))$ have just been exposed to a retweet of $i$'s tweet. Now, for a given retweet burst that occurs between times $t_0$ and $t_1$, we compute:
\begin{align}\label{eqn:predictor}
P(\mbox{follow burst for $i$}\,&|\, \mbox{retweet burst $\in [t_0, t_1)$} ) \nonumber\\ 
\sim &\frac{ \sum_{j\in N_{RT}(i,[t, t+\Delta t))} \hat{P}_{j,i} }{ \sum_{j \in N_2(i) } \hat{P}_{j,i}}.
\end{align}
The above expression simply quantifies the relative fraction of new-follow probability $\hat{P}_{j,i}$ among $i$'s 2-hop neighbors that got exposed to $i$'s retweet.

Now, we can test how well Equation \eqref{eqn:predictor} can predict the occurrence of follow bursts as a means of validating our analysis.  In addition to validation, predicting the appearance of new follow bursts for a given information diffusion event is potentially very useful.  As gaining more followers is an objective of many Twitter users, Equation~\ref{eqn:predictor} can identify where undiscovered new potential followers are in the network, and which of the user's current followers need to retweet her in order for her to obtain new followers.  

\subsection{Testing the model} To quantify how successful the model is at predicting when a retweet burst will cause a follow burst, we devised a simple experiment.  We randomly selected 400K retweet bursts, 21\% of which were followed immediately by a new follow burst for the user.  We were sure not to overlap samples with the ones used to fit Equation~\eqref{eqn:pij}.  Then Equation~\eqref{eqn:predictor} was used to rank the retweet bursts in order of most likely to be succeeded by a new follow burst.  The highest ranked burst was considered our ``first guess'' to be a retweet-follow burst, the second highest ranked burst was our second guess, and so on.  This sequence produced a precision-recall curve, and we calculated the area under the precision-recall curve ($AUC$) as a measure of the performance of the model.  If the model ranks all retweet-follow bursts as most likely, then $AUC = 1$;

\xhdr{Baselines}  We compared the model's performance against a series of baselines.  For each baseline, we used a different property of the retweet burst or a property of the user who is experiencing the burst. Each baseline provides a method of ranking the most likely new follow bursts. For each baseline as well as for our model we compute the area under the precision-recall curve ($AUC$). We consider the following baselines:
\begin{itemize}
\item \textbf{Number of retweet exposures:}  If a user follows someone who retweeted the user as part of the retweet burst, then they have been exposed to the user's tweets.  The retweet bursts are ranked according to the number of such 2-hop neighbors.  The more 2-hop neighbors that are exposed to the user's tweets, the more opportunities for new follows to occur.  As such, this is expected to be the most powerful baseline.
\item \textbf{Number of retweets:}  The retweet bursts are ranked according to the raw number of retweets the user received during the burst.
\item \textbf{Number of followers:} The retweet bursts for users with the largest number of previous follow bursts as ranked first.
\item \textbf{Random:}  The retweet bursts are sorted randomly.
\end{itemize}

\subsection{Results and observations} 

The results of the experiment are shown in Table~\ref{tab:results}.  Our model outperforms each of the baselines by a significant margin, with a $AUC$ score of 0.519 compared to the best baseline score of 0.382.  Ranking based on the number of retweet exposures performed the best out of the baselines, while using the number of followers the user has before the burst did marginally better than random.

\begin{table}
\centering
\begin{tabular}{| l | c |}
\hline
\textbf{Ranking Method} & $AUC$ \\
\hline
Our Model & \textbf{0.518611}\\
Number of retweet exposures & 0.3818 \\
Number of retweets & 0.3340\\
Number of followers & 0.2213\\
Random & 0.2118\\
\hline
\end{tabular}
\caption{Predicting which retweet bursts will cause a subsequent burst in new followers.  Each retweet burst is ranked according to assigned probability of becoming a retweet-follow burst, and the area under the precision-recall curve of the resulting list is calculated. Our model outperforms all the baselines by a significant margin.}
\label{tab:results}
\end{table}

\xhdr{User visibility is less important}  The poor performance of the follower baseline is surprising. In \cite{singer12coevol}, for example, authors found that a user's follower count is highly predictive of future follows.  This may be true in terms of the raw number of new follows a user receives; as we have shown, the number of new follows scales proportionally with increasing follower count.  However, in terms of predicting bursts in new follow arrivals, the current follower count does not perform well.  In some sense, highly retweeted high-degree users are less primed to experience a new follower burst. This is because a large number of potential compatible followers has already been exposed to user's tweets.  This implies that lower degree users are in fact more susceptible to new follower bursts.  In other words, {\em anybody} could potentially experience a burst in new follow arrivals if the circumstances permit.

\begin{figure}
\includegraphics[width=.45\textwidth]{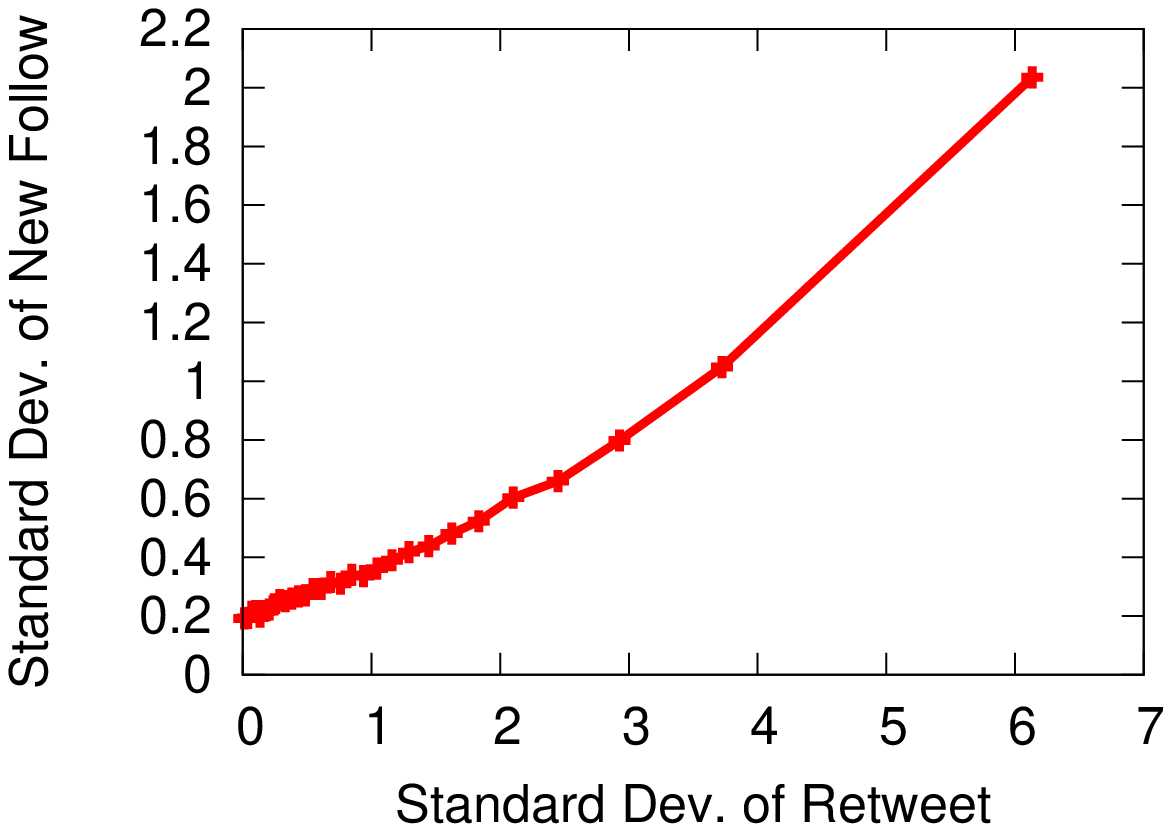}
\caption{We iterated across all high degree users, across every hour of the month in which a single retweet occurred.  The number of standard deviations from the expected value for retweets is plotted against the standard deviations for new follows during the next hour.  
}
\label{fig:rt_standdev_follow_standdev}
\end{figure}

\xhdr{ Retweets are not enough} Using retweets to predict follow bursts is more useful than the indegree of a user.  It is not enough to have a high degree, but rather a user must also participate in information diffusion events in order to experience a new follow burst. The improvement of the Retweet Baseline is also partially due to the fact that the intensity of retweet bursts tends to correlate with the subsequent change in new follows arrivals. Figure~\ref{fig:rt_standdev_follow_standdev} illustrates the phenomena by  correlating the magnitude of the retweet burst (measured in standard deviations from the expected number of retweets) with the magnitude of a follower burst.
We observe that the more intense the retweet burst a user experiences, the more intense the resulting follower burst.

Despite these correlations, the number of retweets experienced during the burst is still not as successful of an indicator as the number of retweet exposures. In other words, while being retweeted frequently might lead to a burst in new follows, it is far more effective to be retweeted by users with large followings themselves.

The main take-away from the success of our model is what conditions are most ideal for the occurrence of a retweet burst.  A user is most likely to experience a burst when her tweets are exposed to a large set of users who are compatible followers (in terms of tweet similarity) but are only discovering her for the first time.  Users who are retweeted regularly tend not to show bursty behavior but instead enjoy a steady-state arrival of new follows.  This also means that the users who are most poised for new follow bursts are ones who have a large portion of nearby similar users that are yet to be discovered.

\section{Related Work}
\label{sec:related}
There have been many works that focus exclusively on the dynamics of networks.  Some of these works have modeled various aspects of network evolution over time \cite{backstrom06groups, yu10strucevol,jure08microevol}.  More recently, research has focused on predicting local changes in the network, such as the addition of specific edges between nodes \cite{backstrom11randomwalk,hutto13followpredict}.  In \cite{kwak12unfollow,xu13brokenties}, the authors predicted the deletion of edges between users.  In the prediction of both edge creation and deletion, many features were found to be useful, including information diffusion-based features. 
Similarly, the diffusion of content through social networks and media continues to be an active field of research \cite{dow2013anatomy,goel2012structure,leskovec-blogspace-sdm07}.  Many works have focused on the different factors that affect the spread of information, including the network structure \cite{kempe03maximizing,romero2013interplay,ugander2012structural}, influence between members of the social network \cite{bakshy2011everyone,bakshy2009social,cha2010measuring}, out-of-network influences such as other forms of media \cite{crane08robust,myers12external}, competing pieces of information \cite{judd-color-2010, myers12clash}, and the nature of the content itself \cite{berger2009social,Hong2011www,kunegis2011bad,tsur2012s}.  
 
Recently, there have been several works that address the effect of information diffusion on network dynamics, and we consider these works to be most closely related to our own. In \cite{weng13infonetevol}, the authors used a mixing of different edge creation null models to find that information diffusion motivates about 12\% of all new edges formed.  Also, \cite{singer12coevol} used autoregression to show how various network and information diffusion properties were correlated in time.  In \cite{antoniades13coevol}, the authors predicted instances of retweet events leading to the formation of new edges.  
Our work, however, focuses on bursts of new follows and unfollows.  We observe that the arrival of new follows as well as unfollows is steady for many users, even those with no tweeting behavior. Moreover, we identify that the interplay between information diffusion and network dynamics lies in the sudden interruption of these steady arrivals.

\section{Conclusion}
\label{sec:conclusion}
In this paper we explored how the burst in network evolution can be a result of information spreading through the network.  We observed burst-like behavior in the network dynamics, as users receive sudden influxes of edge creation or deletion events. We established that such bursts are caused by large information diffusion events. Lastly, we developed a model that can not only predict with high level of accuracy whether a diffusion event will trigger a spike in graph dynamics, but it also provides insight into what causes the occurrence of spikes.  

Potential avenues of future work include a more detailed study of unfollow spikes as well as examining what aspects of user compatibility is important in creating bursts of new followers. Further analysis of the content and the effect of different topics on network dynamics has the potential to lead to new insights. Additionally, extending the model to incorporate the content of each tweet as well would also be interesting.

\xhdr{Acknowledgements}
This research has been supported in part by National Science Foundation grants
CNS-1010921,            
IIS-1016909,            
IIS-1149837,            
IIS-1159679,            
as well as DARPA SMISC,
Albert Yu \& Mary Bechmann Foundation,	
Boeing, 																
Allyes, 																
Samsung,																
Intel,                                  
Alfred P. Sloan Fellowship and 					
the Microsoft Faculty Fellowship. 			
We would also like to thank Twitter, Inc. for giving us access to the data.
\bibliographystyle{abbrv}
\bibliography{refs}

\end{document}